\documentclass[aps,pra,showpacs,twoside,twocolumn,10pt]{revtex4-1}
\usepackage[colorlinks=true, citecolor=red, urlcolor=blue ]{hyperref}
\usepackage{epsfig,newlfont,amssymb,amsfonts,amsmath,bm,subfigure,palatino,mathtools,amsthm,braket,times,soul}
\usepackage[normalem]{ulem}
\usepackage{multirow}

\begin{document}


\title{
Gain in Performance of Teleportation with Uniformity-breaking Distributions}

\author{Saptarshi Roy, Shiladitya Mal and Aditi Sen(De)}

\affiliation{Harish-Chandra Research Institute, HBNI, Chhatnag Road, Jhunsi, Allahabad 211 019, India}

\begin{abstract}
Prior information about the input state can be utilized to enhance the efficiency of quantum teleportation which we quantify using the first two moments of fidelity.
The input knowledge is introduced by relaxing the uniformity assumption in the distribution of the input state and considering non-uniform distributions, namely the polar cap and von Mises-Fisher densities. For these distributions,we show that the average fidelity increases depending on the shared resource state between the sender and the receiver while the deviation decreases with the increase of information content about the input ensemble, thereby establishing its role as a resource. Our comparative study between these two distributions reveals that for the same amount of information content about inputs, although the average fidelity yield is the same for both, the polar cap distribution is “better” as it offers a smaller deviation. Moreover, we contrast the resource of prior information with other resources involved in the protocol like shared entanglement and classical communication. Specifically, we observe that unlike uniform distribution, the amount of classical communication required to fulfill the task decreases with the increase of information available for inputs. We also investigate the role of prior information in higher (three) dimensional teleportation and report the signatures of dimensional advantage in prior information-based teleportation. 

\end{abstract}

\maketitle

\section{Introduction}
The power of quantum physics brings the idea of teleportation from the pages of science fiction to the realms of physical reality \cite{pirandola'15}. From the very inception of quantum teleportation \cite{bennett'93}, it has  always been at the central point of study from various perspectives and  currently turns out to be one of the main ingredients in building  quantum technologies. Apart from instigating the endeavour of quantum information theory, it provides the premise for the art of technology including quantum repeaters \cite{repeater'98}, remote state preparation \cite{bennett'01}, quantum gate teleportation \cite{gottesman'99}, measurement­-based quantum computing \cite{gross'07}, telecloning \cite{murao'99}. 

 In the process of teleportation, some unknown quantum states are transmitted to a distant location without sending the state physically \cite{bennett'93}, which is impossible only by using classical communication. Specifically, non-classicality of teleportation can be quantified by measuring the gap between  the  fidelities obtained by using quantum resources and the one via classical ingredients like the shared  unentangled state \cite{popescu'94}. Fidelity is defined as the overlap between the input  and the  actually teleported states \cite{jozsa'94, horo'96, horo'99}. In the classical protocol, it is computed by some kind of prepare and measure procedure together with  classical communication from the sender, called Alice, to the receiver, Bob \cite{MassarPop} (see  \cite{caval'17} for other notion of nonclassicality in teleportation). 
In laboratories, nonclassical fidelity has successfully been achieved in various physical systems involving photonic systems \cite{bouwm'97, ursin'04, ma'12, yin'12, bin'03}, ion traps \cite{barrett'04, nollek'13}, nuclear magnetic resonances \cite{nmr}, solid state systems \cite{solid'13, gao'13} (see \cite{pirandola'15} for more references).

Teleportation scheme showing quantum advantage typically has several components on which the fidelity depends.  Perfect fidelity is achieved when the shared state is maximally entangled,  Bell measurement at Alice's end  is projective (c.f. \cite{tw}) and the information about the outcomes of the measurement is perfectly communicated to Bob via a noiseless channel. However, this is an ideal situation and all of them never meets with full perfection in  experiments. 
The presence of noise can,  in general, degrade the entanglement content  \cite{entanglementrev} of the  state shared by the sender-receiver duo, thereby lowering the fidelity compared to the one obtained in the perfect case, even when the measurement at Alice's end is perfect. The reduction can sometimes go below the fidelity achieved by  an unentangled state which can be overcome by employing suitable local preprocessing methods developed over the years  on the shared state
\cite{dist1, dist2, dist3, dist4,badziag'00, vers'03,Li'20}.


In the present work, we concentrate on another ingredient, specifically, the distribution of the input state,  involved in teleportation. In this regard,  it was shown that instead of choosing input states uniformly from the entire Bloch sphere if Alice has to teleport one of the two non-orthogonal states, given  randomly with equal probability,   one ebit  (one copy of a maximally entangled state) of the shared resource is necessary which was named as ``two-state teleportation" \cite{henderson'00}.
 Notice that if Alice apriori has the information about the input state, she can teleport only via classical channel without requiring the expensive entangled state while for completely unknown inputs taken uniformly from the Bloch sphere,  perfect teleportation is not possible if Alice-Bob pair share a quantum state having less than one ebit.
 In  another work, optimal teleportation was designed by considering noisy inputs together with a noisy channel \cite{taketani'12}.
 
In our work, we enforce another kind of restrictions on inputs. In particular, instead of limiting the set of states to be teleported,  we impose that the input distribution is not uniform. We show explicitly how prior information about the input states encoded in the non-uniform distribution \emph{enhances} average teleportation fidelity. Specifically, we employ two different distributions over Bloch sphere for choosing inputs, namely polar cap distribution which is obtained by changing the polar angles of the Bloch sphere,  and von Mises-Fisher distribution \cite{Fisher}. 
We characterize the nonclassicality of teleportation protocol via average fidelity where the averaging is performed over the corresponding  distribution of inputs. In addition, we also evaluate  fidelity deviation 
which indicates for a given channel, how different states get teleported far from the target state \cite{bang'18}  (see also \cite{ds, dev1, arka'20, dev2}). 
 In case of the polar cap as well as von Mises-Fisher distributions,  we obtain compact forms of average fidelities  for arbitrary two-qubit density matrices in terms of the correlation matrix  and  show that similar formalisms can also be applied to obtain fidelity deviations.
It is worth mentioning here that the non-uniform distribution  has very recently been considered  in the context of average gate fidelity  to distinguish between two noisy scenarios \cite{nielson'02, gate'20}.
For arbitrary pure two-qubit states, we observe that the average fidelity increases and deviation decreases with the increase of prior information. The  information content of the initial ensemble is quantified by the fidelity of classical cloning i.e., measure-prepare method \cite{MassarPop} in both  situations. Interestingly, we  find that for a fixed  amount of information extractable by using classical protocol, polar cap distribution yields less deviation compared to von Mises-Fisher distribution for pure shared states while the average fidelities for both the distributions coincide. Moreover, we show that in contrast to a  uniform distribution, classical communication (CC) required for obtaining a fixed value of teleportation fidelity  decreases with the increase of information in inputs. Moreover, we present an adaptive local operations and classical communication (LOCC) scheme depending on the prior information required to  enhance the fidelity. Since all the results obtained here are similar to the original teleportation protocol with respect to the measurements   and unitary operations performed by the sender and the receiver respectively,  the proposed scheme  can be experimentally  realized with currently available technologies.

In the last two decades, it was also discovered that in several quantum information processing  tasks,  higher dimensional systems can provide some  advantages over the lower dimensional ones \cite{Cerf'02, highdim3, Durt'08, Vertesi'10, Cava'18, highdim1, highdim2, ZeilingerPan'19}. Towards extracting such power from higher dimensions, we also evaluate the average fidelity of the teleportation protocol for the restricted set of spin-1 inputs and the shared two-qutrit states.  Our analysis confirms that for a fixed value of information contained in inputs,  the gain in fidelity obtained via qutrit systems is approximately double compared to the one that can be achieved via the two-dimensional ones.  

The paper is organized in the following manner. After discussing briefly the method used to evaluate the average teleportation fidelity and fidelity deviation in Sec. \ref{sec:method} which include the way, we quantify, information content in  inputs and the role of adaptive LOCC, we present the results for two choices of distributions for inputs in Sec. \ref{sec:diffdistri}, the polar cap in Subsec. \ref{subsec:polar}, and the von Mises-Fisher distribution in Subsec. \ref{subsec:MF}.  The importance of prior information content in inputs is analysed in Sec. \ref{subsec:utility} in the context of other resources.  The next section (Sec. \ref{sec:comparison}) compares the gain obtained in the  average fidelity and the corresponding deviation by considering the polar cap  and  von Mises-Fisher distribution. In Sec. \ref{sec:highdimen}, we consider the higher-dimensional system and show that the  advantages due to the non-uniform distribution continues. We conclude in Sec. \ref{sec:conclu}.

\section{Methodology}
\label{sec:method}

The success of quantum teleportation \cite{bennett'93} relies on the synergy between various resources from both  quantum and classical domains. The quantum ingredients involved in the protocol are undoubtedly the entanglement of the shared state between Alice and Bob,  the measurement performed by Alice  and the state to be teleported while the required classical one includes the communication of Alice's measurement outcomes to Bob. 
Note that   any components which take part in enhancing the performance of the teleportation protocol are called to be a \emph{resource}.

Here we examine yet another resource, i.e. the \emph{partial information  content of the input state to be teleported.}
Note that we call prior information as resource in the sense that we can utilize it to enhance the teleportation fidelity.
 We will also show the effects of it on other ingredients of the protocol. 

Let us begin by the canonical form of an arbitrary two-qubit state, given by 
\begin{equation}
\rho = \frac{1}{4}(I_4 + \sum_{i=x,y,z} (m_i \sigma_i \otimes I_2 + m'_i I_2 \otimes \sigma_i) + \sum_{i=x,y,z} t_i \sigma_i \otimes \sigma_i),
\end{equation}
where magnetization, \(m_i = \mbox{tr} (\sigma_i \rho_1)\), \(\rho_1\) being the local density matrix at Alice's part,  similarly \(m'_i\), and the correlators are defined as \(t_i = \mbox{tr} (\sigma_i \otimes \sigma_i \rho)\), \(i=x, y, z\) which constitutes a diagonal correlation matrix, $T$. Under the standard teleportation scheme, a two-qubit state with correlation matrix $T$, teleports a state having a Bloch vector $a = (\sin \theta \cos \phi, \sin \theta \sin \phi, \cos \theta)$ with a fidelity of \cite{horo'96}
\begin{eqnarray}
f(\theta, \phi) = \frac12 (1 - a^T T a).
\label{eq:fid1}
\end{eqnarray} 
which reduces  in terms of \(t_i\)s as  
\begin{eqnarray}
f(\theta, \phi) &=& \nonumber \\ \frac12 (1 &-& t_1 \sin^2 \theta \cos^2 \phi - t_2 \sin^2 \theta \sin^2 \phi - t_3 \cos^2 \theta). \nonumber \\ 
\label{eq:fiddis}
\end{eqnarray}
The average fidelity ($F_p$) is obtained by averaging $f(\theta, \phi)$ with respect to a general probability density function on the Bloch sphere, $p(\theta,\phi)$ as
\begin{eqnarray}
F_p = \langle f(\theta,\phi) \rangle = \int_{\theta = 0}^{\pi}  \int_{\phi = 0}^{2\pi} \sin \theta ~p(\theta,\phi) ~f(\theta, \phi) ~d\theta d\phi.\nonumber \\
\label{eq:fidgen}
\end{eqnarray}
The standard deviation of fidelity, commonly known as fidelity deviation \cite{bang'18}, is computed to be
\begin{eqnarray}
D_p = \sqrt{\langle f^2(\theta,\phi) \rangle - F_p^{2}},
\label{eq:dev1}
\end{eqnarray}
where we have
\begin{eqnarray}
\langle f^2(\theta,\phi) \rangle = \int_{\theta = 0}^{\pi}  \int_{\phi = 0}^{2\pi} \sin \theta ~p(\theta,\phi) ~f^2(\theta, \phi) ~d\theta d\phi.
\label{eq:ddd}
\end{eqnarray}
Naturally, $p(\theta, \phi)$ satisfies $\int_{\theta = 0}^{\pi}  \int_{\phi = 0}^{2\pi} \sin \theta ~p(\theta,\phi) ~d\theta d\phi = 1.$
Note that the completely random input case corresponds to a uniform distribution $p^{\text{uni}}(\theta,\phi) = \frac{1}{4\pi}$, for which $F$ for any state with $\det T < 0$  can be expressed as $F^{\text{uni}} = \frac12 (1 + \frac13 \text{ Tr } |T|)$ \cite{horo'96}.
We will discuss how the average fidelity and deviation get altered for different probability densities, $p(\theta, \phi)$s, in the succeeding section. The above method can be extended to a higher dimensional system which will be discussed in Sec. \ref{sec:highdimen} in details.

\subsection{Quantification of prior information}
\label{subsec:quantifyinfo}

Since the idea of our work is to change the distribution of the input states from uniform to a non-uniform one, it is important to quantify the optimal amount of extractable information from an ensemble of states.  When the  set of states  are uniformly distributed in the state space in any dimension $d$, the maximal  information that can be extracted is  given by  \cite{MassarPop}
\begin{eqnarray}
S = \sum_{\xi} \int \mathcal{D}\phi ~P(\phi,\xi) ~S(\xi,\phi),
\end{eqnarray}
where $\xi$ denotes the measurement outcomes, $P(\phi,\xi)$ is the probability of a particular outcome $\xi$ depending on the initial state $\phi$ while $\mathcal{D} \phi$ is the measure for the given distribution of the states $\phi$, and  $S(\xi,\phi)$ denotes the distance between the two states. When $S(\xi,\phi)$ is chosen to be the fidelity distance, $|\langle \xi|\phi\rangle|^2$, $S$  becomes the classical fidelity of teleportation $(F_{cl})$, which is achieved by communicating the measurement outcomes $\xi$ to Bob classically. When the ensemble form a uniform measure on the state space, $S^{uni} = F_{cl}^{uni} = \frac{2}{d+1}$. For a non-uniform ensemble, the maximal amount of extractable information $S = F_{cl}$ is higher than $F_{cl}^{uni}$. We quantify prior information about the input state as the difference between the amount of information that can be obtained from the given ensemble and from  the uniform one, given by
\begin{eqnarray}
I = S - S^{uni} = F_{cl} - F_{cl}^{uni}.
\label{eq:info1}
\end{eqnarray}
One can also consider the fractional enhancement of fidelity as the prior information content,
\begin{eqnarray}
I_f = \frac{F_{cl} - F_{cl}^{uni}}{F_{cl}^{uni}},
\label{eq:info2}
\end{eqnarray}
which turns out to be useful  for inter-dimensional comparisons.

%


\subsection{Adaptive local operations and classical communication protocol}
\label{subsec:adaptiveLOCC}

\begin{figure}
\includegraphics[width=\linewidth]{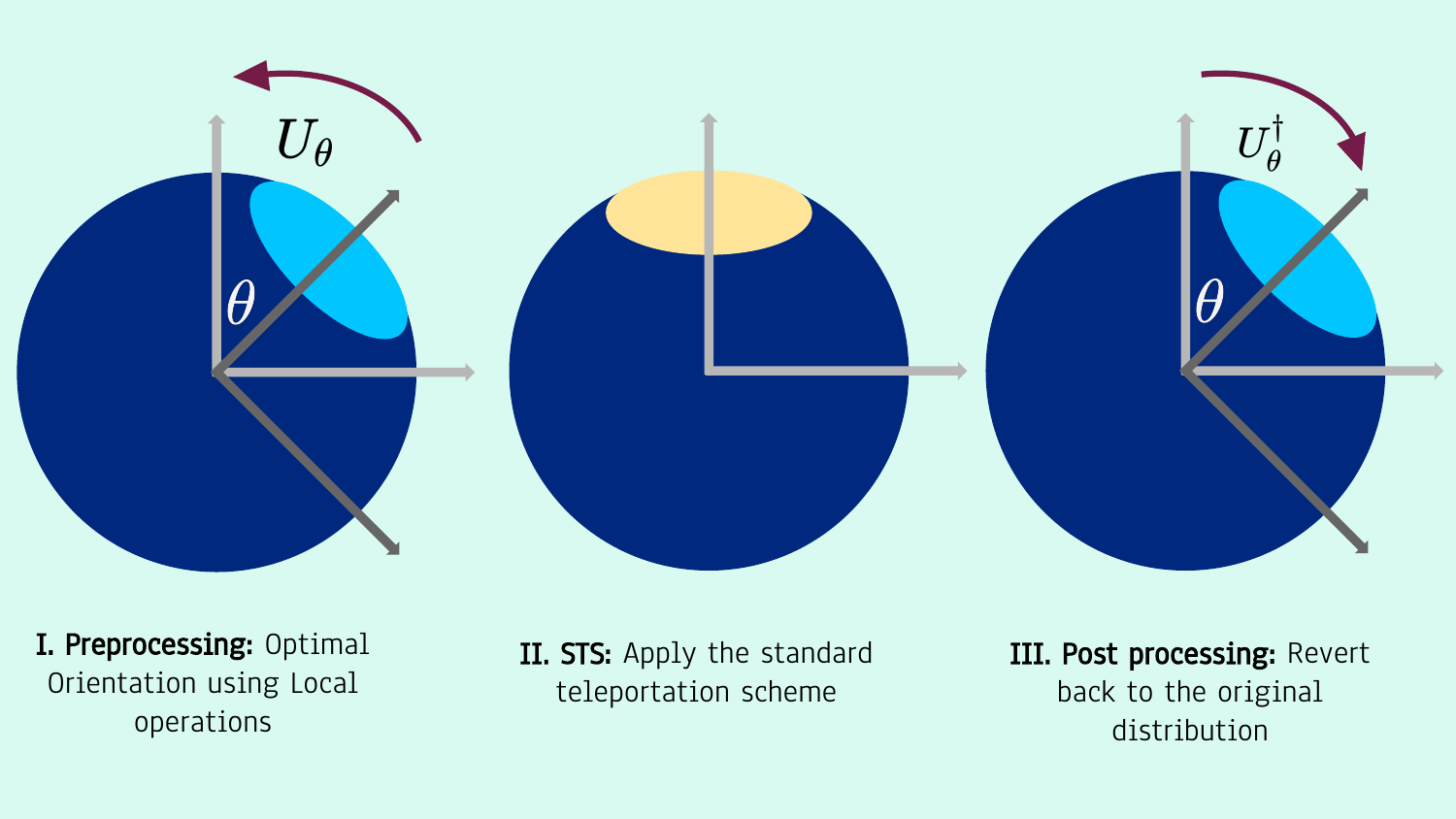}
\caption{LOCC strategy for aligning the probability distribution cap with the pole (measurement direction).}
\label{fig:ap1}
\end{figure}

\begin{figure}
\includegraphics[width=\linewidth]{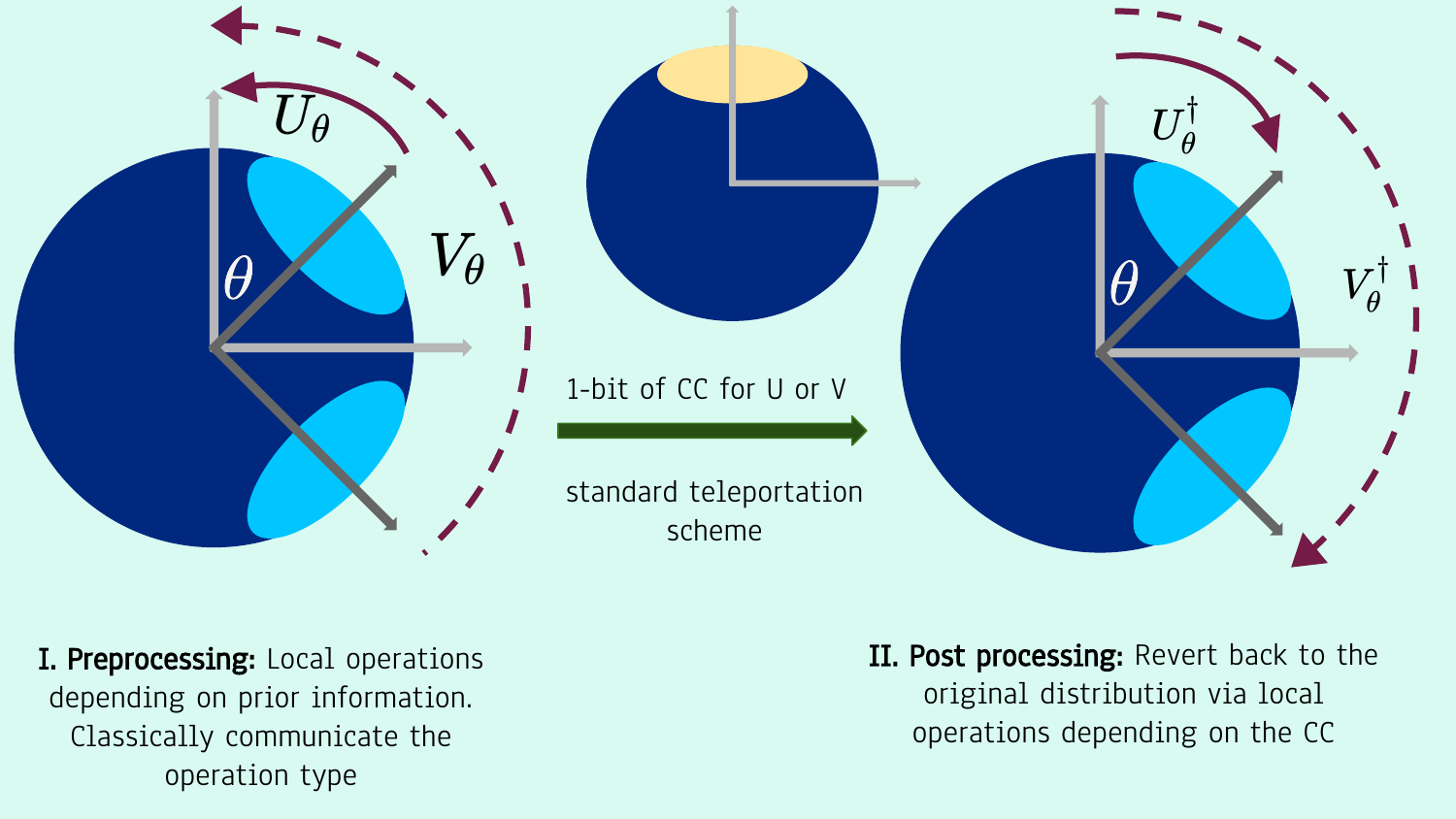}
\caption{LOCC protocol when the input state to be teleported may come from one of two polar cap distributions.}
\label{fig:ap2}
\end{figure}

We now lay out the sketch of  the general strategy to implement an adaptive local operations and classical communication protocol on top of the standard teleportation scheme (STS) to maximize the fidelity output depending on the prior information. The general LOCC scheme has the following two parts:
\begin{enumerate}
\item \emph{Pre-processing.} Depending on the partial information of the distribution from which the states to be teleported come from, before starting the protocol, Alice performs one of the $M$ local operations  to orient the input state depending on the symmetries of the shared state and the distribution for the optimal fidelity yield via the STS as shown in Figs. \ref{fig:ap1} and \ref{fig:ap2} for the particular case of inputs coming from a polar cap on the Bloch sphere. It would be discussed in detail in subsequent sections.
\item \emph{Post-processing.} Depending on the  classical communication of  (CC) $\log_2 M$-bits of data to Bob  about measurement outcomes by Alice, 
Bob re-orients the teleported state back to the original distribution (for post-processing in the particular case of polar distribution of inputs, see Fig. \ref{fig:ap2}).
\end{enumerate}

Note, however,  that such an LOCC update of standard teleportation protocol depends on the partial knowledge of the input as well as on the shared entangled state. For example, for the same polar cap distribution, if the shared state is a Werner state,  there is no point in updating the protocol since the Werner states teleport all states of the Bloch sphere with identical fidelities (since it possesses zero deviation). We shall elaborate on this point in the subsequent sections. We must also add that without applying a LOCC protocol, not only the obtained fidelity in presence of prior information can fall below the best classical (entanglement-free) protocol but can  even be lower than the classical fidelity ($2/3$) in the uniform case.

\section{Effects of prior information  on fidelity}
\label{sec:diffdistri}

We can expect that when some information about the input state to be teleported is available,  the relevant parties can use that information to enhance the performance of the teleportation which can be measured by the fidelity or the fidelity deviation. As we will show, such an enhancement is not ubiquitous and is dependent on the shared entangled resource. In this section, we perform a detailed analysis of how some prior information about the input state in terms of distributions effects the teleportation fidelity. The knowledge about the input comes from the fact that there can be situations where the state to be teleported comes from a specific region of the Bloch sphere. More generally, some states on the Bloch sphere may be more probable to get teleported than  others.
We model such prior information as non-uniform probability density functions on the Bloch sphere and illustrate via two scenarios -- (1) input is given from the polar cap and (2) from the von Mises-Fisher distribution. 

\subsection{The polar cap distribution}
\label{subsec:polar}

\begin{figure}[h]
\includegraphics[width=0.95\linewidth]{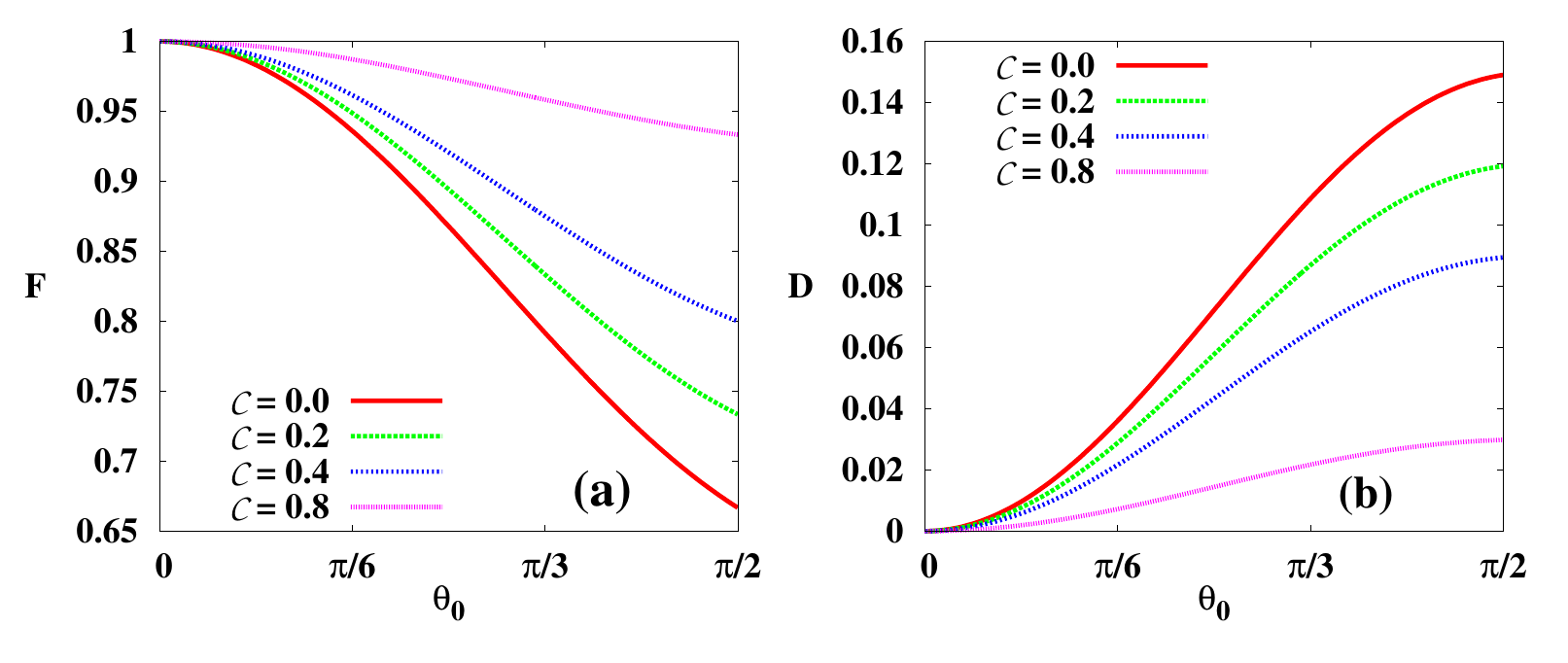}
\caption{(a) Average fidelities (ordinate) and (b) their deviations  (ordinate) vs.  the various sizes of the polar cap as controlled by $\theta_0$ (abscissa). Different curves correspond to the different contents of entanglement of the  shared pure state, quantified by concurrence. The curve with the legend $\mathcal{C} = 0$ represents a shared unentangled state between the sender and the receiver for which the classical protocol has to be followed. 
Note that smaller the polar cap, as indicated by lower $\theta_0$ values, greater is the obtained fidelity and lower is the deviation. Interestingly, for $\theta_0 \to 0$,  the states at or near the pole always get teleported with almost unit fidelity for any values of $\mathcal{C}$ which can be clearly seen from Eq. \eqref{eq:fidalpha}. Both the axes are dimensionless. }
\label{fig:polar}
\end{figure}

Let us start with the situation where the input state is chosen from a polar cap of the Bloch sphere which is obtained by uniformly  varying the polar angle of the sphere.
In this case, $p(\theta,\phi)$ assumes the following form:
\begin{eqnarray}
p_{\theta_0}(\theta, \phi) = \left\{
 \begin{array}{cc}
 \frac{1}{2\pi(1-\cos \theta_0)}, &\text{ for } \theta \leq \theta_0  \\
 0, & \text{ for } \theta > \theta_0
\end{array}\right..
\label{eq:polar}
\end{eqnarray} 
Note that the above probability density is normalized $\int_{\theta = 0}^{\theta_0} \int_{\phi = 0}^{2\pi} p_{\theta_0} \sin \theta d\theta d\phi  = 1.$ It directly follows from the fact that $\int_{\theta = 0}^{\theta_0} \int_{\phi = 0}^{2\pi} \sin \theta d\theta d\phi  = 2\pi(1 - \cos \theta_0)$ and $ p_{\theta_0} = \frac{1}{ 2\pi(1 - \cos \theta_0)}.$
Furthermore, notice  that $\theta_0$ controls the size of the polar cap. Mathematically, it limits the range of the integration for obtaining the average fidelity and deviation in Eqs. \eqref{eq:fidgen} and \eqref{eq:ddd}.

The average fidelity for a given polar cap distribution from Eq. \eqref{eq:fidgen} for an arbitrary two-qubit state can be computed and is given by 
\begin{eqnarray}
F(\theta_0) = \frac12 \big[ 1 &-& \frac{(2+\cos\theta_0)(1-\cos \theta_0)}{6}(t_1 + t_2) \nonumber \\
 &-& \frac{t_3}{3}(1 + \cos\theta_0 + \cos^2\theta_0) \big].
\label{eq:fidgenpolar}
\end{eqnarray}
In a similar fashion, we can obtain  
the fidelity deviation which we are not presenting here since it looks quite cumbersome. 

Let us consider here three classes of shared states, arbitrary pure states where enhancement of fidelity,  its deviation due to prior information is possible and the Bell diagonal (BD) states with different ranks and a mixture of nonmaximally entangled states with product ones.  We show that for the second case, such increment in fidelity by a standard teleportation protocol  is not always possible while the third one yields enhanced fidelity.

\emph{Pure states.} An arbitrary two-qubit pure shared state in Schmidt decomposition can be written as 
\begin{eqnarray}
|\psi_\alpha\rangle = \sqrt{\alpha}|01\rangle - \sqrt{1-\alpha}|10\rangle.
\label{eq:psialpha}
\end{eqnarray}
Note that $|\psi_\alpha\rangle$ possesses the following correlation matrix, $T_{\alpha} = -\text{diag} (\mathcal{C}_\alpha,\mathcal{C}_\alpha,1)$, where $\mathcal{C}_\alpha = 2\sqrt{\alpha(1-\alpha)}$ is the concurrence \cite{Wootterseof} of the same. In this context, the expression of fidelity as in Eq. \eqref{eq:fid1} is computed to be
\begin{eqnarray}
f(\theta, \phi) = 1 - \frac12 \sin^2 \theta (1-\mathcal{C}_\alpha),
\end{eqnarray}
and consequently, we get
\begin{eqnarray}
f^2(\theta, \phi) = 1 + \frac14 \sin^4 \theta (1-\mathcal{C}_\alpha)^2 - \sin^2 \theta (1-\mathcal{C}_\alpha).
\end{eqnarray}
The average fidelity with the Bloch sphere weight as given in Eq. \eqref{eq:polar} turns out to be
\begin{eqnarray}
F(\theta_0) = 1 - \frac{1}{6} \big[(1-\mathcal{C}_\alpha)(2+ \cos \theta_0) (1-\cos \theta_0)\big], \nonumber \\ 
\label{eq:fidalpha}
\end{eqnarray}
while the corresponding fidelity deviation is given by
\begin{eqnarray}
D(\theta_0) = \frac{1-\mathcal{C}_{\alpha}}{6\sqrt{5}}(1-\cos \theta_0)\sqrt{4\cos^2 \theta_0 + 7\cos \theta_0 + 4}. \nonumber \\
\end{eqnarray}
Note that  the uniform case can be recovered with $F(\pi) = \frac{2+\mathcal{C}_\alpha}{3}$ and $D(\pi) = \frac{1-\mathcal{C}_{\alpha}}{3\sqrt{5}} =  \frac{1-F(\pi)}{3\sqrt{5}}$ \cite{arka'20}.
With the increase of  \(\theta_0\), the fidelity decreases for a fixed amount of entanglement while the opposite picture emerges for  deviations as depicted in Fig. \ref{fig:polar}. Speciifically, high entangled shared state leads to less decrease in fidelity with the prior information of polar angle compared to that of the shared state having low entanglement.  Furthermore, note that for a given polar cap extent $\theta_0$, the fidelity yields in the classical scheme \cite{MassarPop}  as 
\begin{eqnarray}
F_{cl}(\theta_0) = 1 - \frac{1}{6} \big[(2+ \cos \theta_0) (1-\cos \theta_0)\big].
\label{eq:fidclassical}
\end{eqnarray}
Therefore, for a given $\theta_0$, any fidelity, $F$ can be called non-classical if and only if $F > F_{cl}(\theta_0)$. Interestingly, for $\theta_0 \to 0$,  the states at or near the pole always get teleported with almost unit fidelity for any values of $\mathcal{C}$ which can be clearly seen from Eq. \eqref{eq:fidalpha}. This can be physically understood by noting that for $\theta_0 \to 0$, we essentially want to teleport either $|0\rangle$ or $|1\rangle$ for which only $1$-bit of CC is sufficient and no entanglement is required. It is reflected in the fact $F_{cl}(\theta_0 \to 0) \to 1$ in this limit as well.


\emph{Werner states \cite{Werner'89}.} Next we repeat the analysis for Werner states, which is given by
\begin{eqnarray}
\rho_W = p |\psi^-\rangle\langle \psi^-|+ \frac{1-p}{4}\mathbb{I}_4,
\end{eqnarray} 
where $|\psi^-\rangle = \frac{1}{\sqrt{2}}(|01\rangle - |10\rangle)$, and $0 \leq p \leq 1$. Note that $\rho_W$ is entangled and consequently yield nonclassical teleportation fidelity for $p>1/3$ in the uniform case.  The  correlation matrix,  of $\rho_W$ can be expressed as $T_{p} = -p~\text{diag} (1,1,1)$. The expression of fidelity as in Eq. \eqref{eq:fid1} is independent of the input state and is computed to be
\begin{eqnarray}
f(\theta, \phi) = \frac{1+p}{2}.
\label{eq:fidW1}
\end{eqnarray}
Therefore, from Eq. \eqref{eq:fidgen}, it is clear that $f(\theta, \phi)$ comes out of the integral as a constant and the normalized $p(\theta, \phi)$ integrates to unity. Consequently, the average fidelity turns out to be independent of the probability density function, given by
\begin{eqnarray}
F(\rho_W) = \frac{1+ p}{2}.
\label{eq:fidW2}
\end{eqnarray}
Therefore, one \emph{cannot} enhance the teleportation fidelity with the prior knowledge about the input. It is a direct consequence of the fact that  the fidelity deviation for $\rho_W$ also vanishes identically. It follows straightforwardly from Eqs. \eqref{eq:dev1}, \eqref{eq:fidW1}, and \eqref{eq:fidW2}. 

Although in the case of uniformly distributed inputs, Werner states turn out to be the one which yield the maximal fidelity for a given concurrence, the case is different when one considers the input states chosen from the polar cap. Since the fidelity enhancement is not possible, Werner states fall behind from their optimal status. In particular, despite being entangled,  it yields a non-classical fidelity only when $F(\rho_W) > F_{cl}(\theta_0)$, i.e.,
\begin{eqnarray}
p > 1 - \frac{1}{3} \big[(2 + \cos \theta_0) (1-\cos \theta_0)\big] = p^*(\theta_0).
\end{eqnarray}
Therefore, when $p \in (1/3, p^*(\theta_0)]$, the Werner states are entangled but does not provide a nonclassical teleportation fidelity. For example, $p^*(\theta_0= \pi/3) = 7/12 = 1/3 + 1/4$. Equivalently, in terms of entanglement, the Werner states provide quantum advantage when their entanglement is greater than $\frac{3p^*(\theta_0) - 1}{2}$, which for $\theta_0 = \pi/3$ turns out to be $3/8$.

\emph{Bell diagonal states.} To better understand the situation, let us now move to the example of Bell diagonal (BD) states, \(\rho = \sum_i p_i |\psi_i\rangle \langle \psi_i|\), where \(|\psi_i\rangle\)s are Bell states \cite{BD}. We segregate the  BD states into two ways, namely via rank, and presence (or absence) of axial symmetry. Mathematically, in our convention, axial symmetry is reflected in the first two entries of the correlation matrix,  i.e.,  when $t_{1}$ and $t_{2}$ are identical. Our earlier examples of the non-maximally entangled pure states and the Werner states possess the axial symmetry. Moreover, we note that rank-2 BD states are always axially symmetric and their teleportation properties are same as the non-maximally entangled pure states in Eq. \eqref{eq:psialpha}  for a fixed value of the entanglement content. It can be easily deciphered from the fact that both of their correlation matrices are identical, and of the form $-\text{diag} ~\lbrace \mathcal{C}, \mathcal{C}, 1 \rbrace$, where $\mathcal{C}$ is the entanglement (concurrence) of the shared state. Note that the concurrence of any BD state can be expressed in terms of the maximal mixing probability $p$ as $\mathcal{C} = 2p - 1$.

On the other hand,  BD states having rank-3 do not possess any axial symmetry. If the mixing probabilities are $p, q$, and $1-p-q$, with $p$ being the maximal weight, the corresponding correlation matrix is $-\text{diag} \lbrace 2p -1, 1-2q, 2(p+q) -1 \rbrace$.
Although one can use the correlation matrix elements to straightforwardly compute the average fidelity using Eq. \eqref{eq:fidgenpolar},  for purposes of illustration, we choose $q = \frac{1-p}{2}$ which provides some mathematical simplifications without compromising on the physical insights. For a given, $\theta_0$, the critical value of $p$ for this class of BD states reads as
\begin{eqnarray}
p > p^*(\theta_0) = \frac{4 + \cos \theta_0 + \cos \theta_0^2}{8 - \cos \theta_0 - \cos \theta_0^2}.
\end{eqnarray} 
For $\theta_0 = \pi/3$, we get $p^*(\theta_0 = \pi/3) = 19/29$ and the corresponding critical entanglement is $2p^*(\theta_0 = \pi/3) - 1 = 9/29$, thereby providing a range of $p$ for which the entangled BD states are not useful for standard teleportation when inputs are chosen from the polar cap distribution. Recall the corresponding critical entanglement value for the Werner states was $3/8 > 9/29$. Similar examples can also be found from the sector of  BD states with rank-4 which houses both axially symmetric states and ones which do not possess such symmetry. Unlike arbitrary rank-1 states and rank-2 BD states,  these examples show that  prior information contained in the polar cap distribution does not enhance the standard teleportation fidelity in case of higher rank BD states since a finite gap opens up where a critical value of non-zero entanglement is required to extract a non-classical value of average fidelity.

We now present an explanation why prior information (in terms of polar cap distribution of inputs) in the case for BD states of rank $> 2$ is not enough for obtaining quantum advantage. In particular, we take up an example of a rank-$4$ axially symmetric BD state for our purpose of demonstration. To understand this, let us briefly recall the reason behind  the quantum advantage  for certain other examples considered before, when the inputs come from a polar cap. It simply follows from  Eq. \eqref{eq:fidgenpolar} that for those shared resource states, input states chosen near the poles are teleported with higher fidelities compared to inputs near the equator of the Bloch sphere. The enhancement relies precisely on this fact of symmetry matching where the input distribution coincides with the region of the Bloch sphere for which the shared resource state teleports states with better than average fidelities.

Consider the following rank-$4$ axially symmetric BD states with mixing probabilities ($p_i$s) being $\{p,q,q,r\}$ and having the following diagonal correlators $t_1 = t_2 = -(p-r)$, and $t_3 = -(p -2q + r)$. Now, from Eq. \eqref{eq:fiddis}, we can clearly see that the chosen class of BD states teleports states near the equator with higher than average fidelity while the states near the poles are teleported with fidelity that falls below the average value. Now, when we contemplate this feature in the context of the polar cap distribution, it reveals a clear mismatch of regions from where the input states are chosen for teleportation and for which the shared state teleports with above average fidelity.  
However, this does not mean that these BD states are not useful. If the inputs to be teleported come from a region that coincides with the high fidelity region of the shared state, these states can  show quantum advantage. An example of such a distribution  takes the form of an annular strip around the equator, and can be defined via the following probability density function:
\begin{eqnarray}
\bar{p}_{\theta_0}(\theta, \phi) = \left\{
 \begin{array}{cc}
 \frac{1}{4\pi\sin \theta_0}, &\text{ for } \frac{\pi}{2}-\theta_0 \leq  \theta \leq \frac{\pi}{2}+\theta_0  \\
 0, & \text{ otherwise } 
\end{array}\right.,
\label{eq:annular}
\end{eqnarray} for $\theta_0 \leq \frac{\pi}{2}$. For such a distribution, one can always fetch a non-zero quantum advantage with prior information for the considered class of rank-$4$ BD states provided $p>\frac12$ and $\theta_0 < \frac{\pi}{2}$.

We now present an example of a mixed state that yields nonclassical fidelities in presence of prior information in the form of the polar cap distribution. It reads as
\begin{eqnarray}
\rho_p = p|\psi_{\alpha}\rangle\langle\psi_{\alpha}| + \frac{1-p}{2}(|01\rangle\langle01| + |10\rangle\langle10|).
\end{eqnarray}
Note that $\rho_p$ is entangled for $p\mathcal{C}_{\alpha}>0$ with its entanglement as measured by concurrence is given by $\mathcal{C}(\rho_p) = p \mathcal{C}_\alpha$. The corresponding diagonal correlation matrix elements of $\rho_p$ are 
\begin{eqnarray}
t_1 = t_2 = -p \mathcal{C}_\alpha, \text{ and } ~t_3 = -1.
\end{eqnarray}
Note that for $\mathcal{C}_\alpha = 1$, $\rho_p$ reduces to a rank-2 BD state. For a polar cap with a cap extent of $\theta_0$, the average fidelity $F_p(\theta_0)$ is calculated using Eq. \eqref{eq:fidgenpolar}, and for $p = 0$, we get $F_{p=0}(\theta_0) = F_{cl}(\theta_0)$. For any $p>0$, we get nonclassical fidelities when the states to be teleported come from the polar cap distribution, with $F_{p>0}(\theta_0) > F_{cl}(\theta_0)$.



\subsubsection{Adaptive LOCC protocol for non-maximally entangled states}

Let us illustrate the LOCC strategy mentioned in SubSec. \ref{subsec:adaptiveLOCC}  when the shared entangled state is $|\psi_{\alpha}\rangle = \sqrt{\alpha}|01\rangle - \sqrt{1-\alpha}|10\rangle$, and the input states to be teleported come from a polar cap distribution. Now the pole of the polar cap might be in a different direction compared to the basis in which the shared state is prepared, and   without any loss of generality, we assume to be the same as the basis in which the Bell measurements are performed. The basis alignment is achieved by applying an unitary $U_{\theta}$ as depicted in Fig. \ref{fig:ap1}.  Since the fidelity distribution of $|\psi_{\alpha}\rangle = \sqrt{\alpha}|01\rangle - \sqrt{1-\alpha}|10\rangle$ is axially symmetric,  and the angular difference between the basis for $|\psi_{\alpha}\rangle$ and the polar cap is $\theta$, the receiver, in addition, in implementing the standard Pauli unitaries depending on the outcomes of the initial Bell measurements, must also act with $U_{\theta}^{\dagger}$, which will rotate the teleported states appropriately. So in this case, the input knowledge updates the final unitaries as $U_i = U_{\theta}^{\dagger}\sigma_i$, see Fig. \ref{fig:ap1}.

We now present a more general example where the requirement for an adaptive LOCC protocol becomes more involved from a basis orientation. Suppose instead of a single cap, the state to be teleported comes from either of the two caps as depicted in Fig. \ref{fig:ap2}. At the time when the state to be teleported is handed over to Alice, the information about  the two caps  is also provided. Depending on this information, Alice first applies $U_{\theta} (V_{\theta})$ which aligns the cap(s) with the pole, and then proceeds with the standard teleportation scheme. Additionally, Alice sends Bob one additional bit of information informing the particular cap from which the teleported state came from. After completing the STS,  Bob, depending on that additional 1-bit of CC by Alice, applies $U_{\theta}^{\dagger} (V_{\theta}^{\dagger})$, thereby correctly recovering the initial location (cap 1 or cap 2) from which the state for teleportation came from. It constitutes the adaptive LOCC protocol for maximizing the teleportation fidelity in presence of prior information. Notably, if more than two, say $d$ non-overlapping caps are present, we can simply generalize our strategy using d encoding and decoding unitaries and an additional $\log_2d$-bits of CC from Alice to Bob.    See Fig. \ref{fig:ap2} for a schematic representation.

\subsection{The von Mises-Fisher distribution for inputs}
\label{subsec:MF}

We now move on to the situation when the probability of the input state follows a von Mises-Fisher distribution \cite{Fisher} over the Bloch sphere which can be expressed as
\begin{eqnarray}
p_{\kappa}(\theta, \phi) = \frac{\kappa}{4\pi \sinh \kappa} ~\exp ~(\kappa \cos \theta),
\label{eq:fisher}
\end{eqnarray}
where $\kappa \geq 0$ is the concentration parameter. For $\kappa = 0$, we have $p(\theta, \phi) \sim$ constant corresponding to an uniform distribution over the Bloch sphere. When $\kappa$ is large, the distribution is concentrated near the pole with $\theta = 0$. Note that the distribution is symmetric or can be called uniform with respect to the azimuthal angle $\phi$. It shares some statistical similarities with the Gaussian distribution, which can be obtained for large $\kappa$, see \cite{book}.

Using  Eq. \eqref{eq:fidgen}, the average fidelity of an arbitrary two-qubit state for  this distribution in terms of  correlators turns out to be
\begin{eqnarray}
F(\kappa) = \frac12 \big[ 1 &-& \frac{(t_1 + t_2)}{\kappa^2}(\kappa \coth \kappa - 1)  \nonumber\\
 &-& \frac{t_3}{\kappa^2}\big\lbrace(2+\kappa^2) - 2\kappa \coth \kappa \big\rbrace \big].
 \label{eq:test}
\end{eqnarray}
The corresponding fidelity deviation can be obtained from Eq. \eqref{eq:dev1} with
\begin{widetext}
\begin{eqnarray}
\langle f^2 \rangle &=& \frac{1}{4 \kappa^4}[\kappa^4 (t_3-1)^2 + 3 (3 t_1^2 + 2 t_1t_2 + 3 t_2^2 - 8 (t_1 + t_2) t_3 + 8 t_3^2) \nonumber \\ 
&+& \kappa^2 (3 t_1^2 + t_2 (2 + 3 t_2) + 2 t_1 (1 + t_2 - 5 t_3) - 2 (2 + 5 t_2) t_3 + 12 t_3^2) \nonumber \\ 
&+&  \kappa (2 \kappa^2 (t_1 + t_2 - 2 t_3) (t_3 - 1) - 3 (3 t_1^2 + 2 t_1 t_2 + 3 t_2^2 - 8 (t_1 + t_2) t_3 + 8 t_3^2)) \coth \kappa], \nonumber
\end{eqnarray}
\end{widetext}
while $F (\kappa)$ being computed via  Eq. \eqref{eq:test}.

 To demonstrate the effects of prior information of inputs via von Mises-Fisher distribution, we investigate the behavior of fidelity and its deviation by fixing entanglement of the shared state.   

\begin{figure}[ht]
\includegraphics[width=0.95\linewidth]{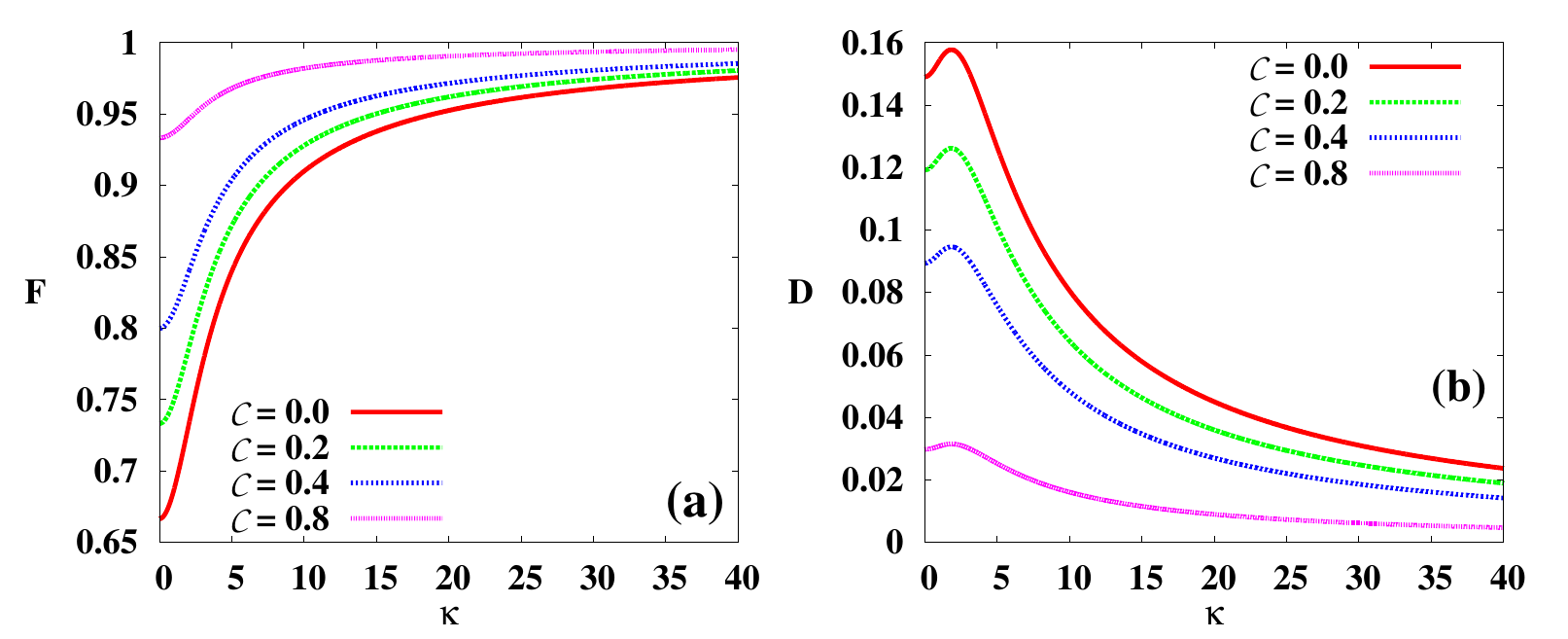}
\caption{(a) Average fidelities  (vertical axis) for the von Mises-Fisher distribution with different concentration parameters,  $\kappa \in (0,10]$ (in the horizontal axis). (b) Trends of fidelity deviations with respect to \(\kappa\). Different curves are for various chosen fixed values of concurrence, \(C_{\alpha}\) of an arbitrary pure states. Note that a high value of $\kappa$ indicates a more concentrated distribution around the north-pole of the Bloch sphere, while $\kappa = 0$ corresponds to an uniform average. Therefore, we get a higher value of average fidelity and lower deviation for large $\kappa$ compared to the case with low \(\kappa\).   Again, as  seen in  Fig. \ref{fig:polar}, when states to be teleported  are chosen at or very near the pole, which in this case is the large $\kappa$ limit, the output states produced are close to the input ones, thereby giving close to the  unit fidelity. Both the axes are dimensionless.}
\label{fig:vMF}
\end{figure}

\emph{Example 1: Pure states.} The average fidelity for an arbitrary pure state,  $|\psi_{\alpha}\rangle$ in Eq. \eqref{eq:psialpha} in this case simplifies as
\begin{eqnarray}
F(\kappa) = 1 + \frac{1-\mathcal{C}_{\alpha}}{\kappa^2}(1 -\kappa \coth \kappa),
\label{eq:fidpsifisher}
\end{eqnarray}
while its deviation reads as
\begin{eqnarray}
D(\kappa) = \frac{1-\mathcal{C}_{\alpha}}{\kappa^2}\sqrt{2\kappa^2 + 6(1-\kappa\coth \kappa)-(1-\kappa\coth \kappa)^2}.\nonumber\\
\end{eqnarray}
Note that the classical protocol  with the shared unentangled states yields a fidelity of 
\begin{eqnarray}
F_{cl}(\kappa) = 1 + \frac{1}{\kappa^2}(1 -\kappa \coth \kappa).
\label{eq:fidclassicalfisher}
\end{eqnarray}
In this case also, the pattern for average fidelity and deviation with the increase of prior information content shows similar behavior as seen for the polar cap. Note that the apparently opposite trend  observed between Figs.  \ref{fig:polar}  and \ref{fig:vMF}  is due to the fact that the information content is high when \(\kappa\) is high and when \(\theta_0\) is low for the  polar cap distribution.

\emph{Example 2: Werner states.}
An entangled Werner state yields average fidelity above the classical limit when $F(\rho_W) > F_{cl}(\kappa)$. Applications of   Eqs. \eqref{eq:fidW2} and \eqref{eq:fidclassicalfisher} lead to the  bound on the mixing parameter, given by
\begin{eqnarray}
p > 1 + \frac{2(1-\kappa \coth \kappa)}{\kappa^2} = p^*(\kappa).
\end{eqnarray} 
Although,  the Werner states are entangled with $p \in (1/3, p^*(\kappa)]$, states do not give any quantum advantage by using the standard teleportation protocol. For example, $p^*(\kappa = 10) = 0.679 = 1/3 + 0.346$. In terms of entanglement, the Werner states gives quantum fidelities when their entanglement is greater than $\frac{3p^*(\kappa) - 1}{2}$, which for $\kappa = 10$ is computed to be $0.518$. Similar analysis can be carried out for the BD states as well, and expectedly, the results turn out to be qualitatively similar with those obtained  for the polar cap distribution, i.e., with the increase of rank, or with the violation of axial symmetry, the BD states develop a gap between entanglement and teleportability, thereby illustrating a requirement for designing a teleportation scheme beyond the standard one. 

The enhancing features based on presence of prior knowledge about the input state of teleportation can be summarized in two requirements:
\begin{enumerate}
\item The shared state needs to be asymmetric in the sense that different states of the Bloch sphere are teleported with non-identical fidelities, i.e., it possesses non-zero fidelity deviation.                                            
\item The shape of the non-uniform distribution should be such that it can pick out the states to be teleported with higher than average fidelity.
\end{enumerate}. 
These two conditions must be met to obtain any fidelity enhancement depending on the prior knowledge. Therefore, apart from the resource state and distribution combinations we have stated as examples, there can be many other combinations that provide prior information induced quantum advantage.

\begin{figure}[ht]
\includegraphics[width=\linewidth]{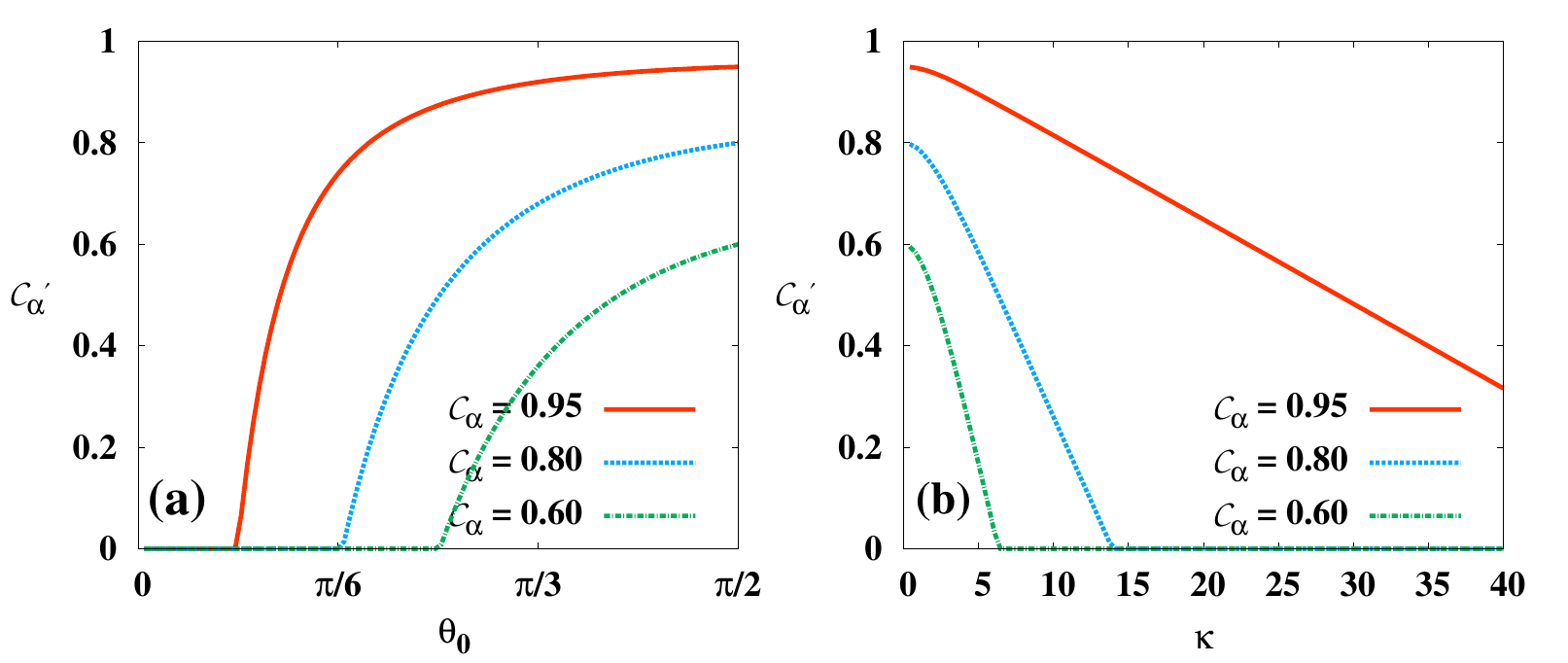}
\caption{Entanglement required  (ordinate) to reproduce a given fidelity in the uniform case when inputs are chosen from the polar cap in (a) and von Mises-Fisher density  in (b) vs. the corresponding parameters (abscissae).  Different curves correspond to a fixed amount of  entanglement of the shared  pure states which also indicate the fidelity achieved by the uniform distributions of inputs. Interestingly, we notice that when   content of information available for inputs is high, the classical protocol itself can reproduce the required fidelity without the need of any supplementary entanglement. The vertical axis is in ebits and the horizontal axis is dimensionless. }
\label{fig:resource}
\end{figure}

\begin{figure}[ht]
\includegraphics[width=\linewidth]{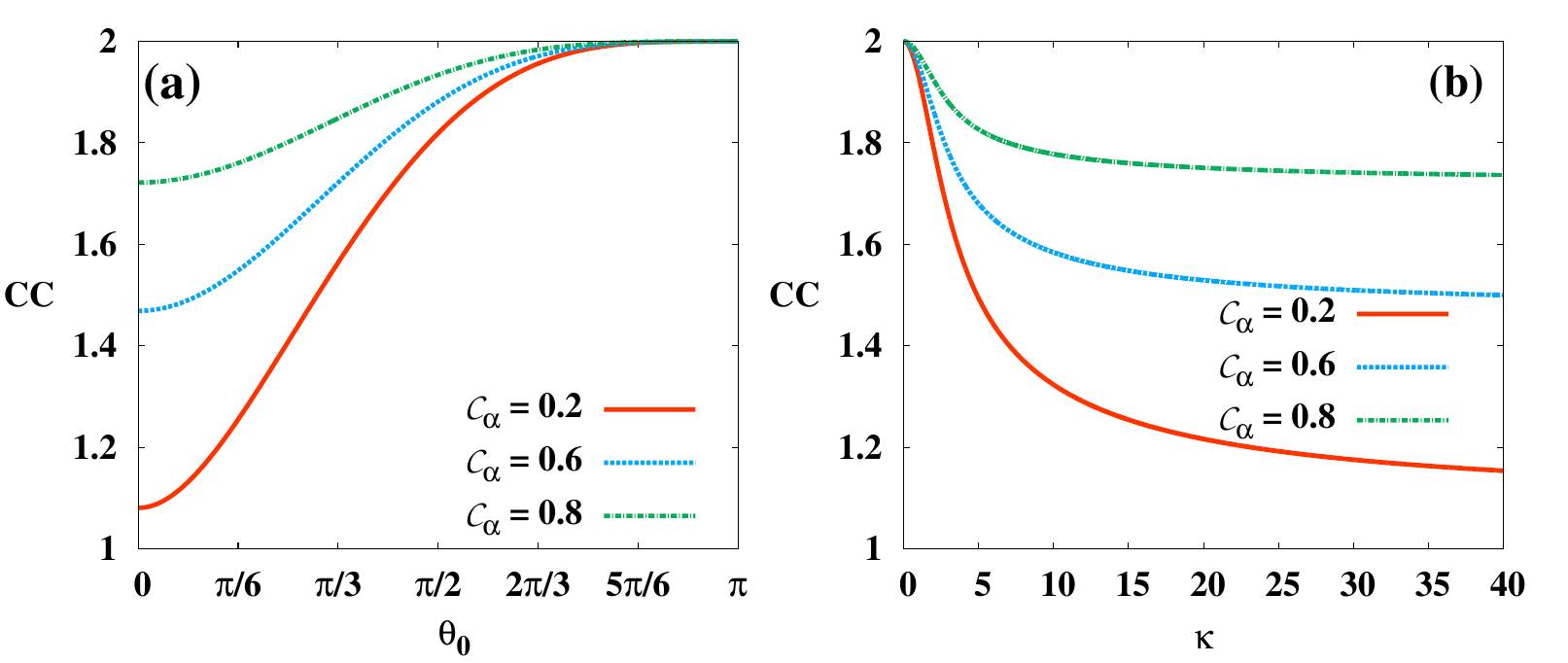}
\caption{The amount of classical communication (vertical axis) required for faithful transmission of  information about Bell measurement when input states are chosen from (a) polar cap, and (b) von Mises-Fisher distribution   for different values of entanglement of the shared  pure states. CC is quantified via Shannon entropy of the probabilities of the outcomes of Bell measurement. The vertical axis is in bits while the horizontal axis is dimensionless.  } 
\label{fig:resource-cc}
\end{figure}

\subsection{Utility of prior information in resource reduction}
\label{subsec:utility}

Let us ask a simple question: \emph{Can knowledge about the input state  be considered as a resource?} 
To address this question, we take a step back and ask why entanglement is considered as a resource in quantum teleportation. The answer is simple. Without entanglement, one cannot obtain a nonclassical fidelity, and the maximal fidelity obtained for a given entanglement value grows monotonically as  entanglement increases, with perfect teleportation being achieved for the maximally entangled one. 
 In the previous section, we have already shown that  information about the distribution of the input state can enhance the performance of the teleportation protocol, thereby establishing prior  information about the choice of distribution of the input as resource.  Let us now compare the trade-off between different resources involved in this scheme.  Specifically, we study the way, \emph{prior information about the input state} can alter the consumption of other resources like shared entanglement and classical communication cost ($2$-bits) during the standard protocol.

\subsubsection{Prior information vs. entanglement} 
\label{subsec:infovsent}

To establish a connection between the shared entanglement required for an uniform distribution and for the  distributions considered in this work, let us take an arbitrary pure state, $|\psi_{\alpha}\rangle$  in Eq. \eqref{eq:psialpha}. For a given $\alpha$, for a completely unknown input chosen from the uniform distribution, we obtain an average fidelity of $ \frac{2 + \mathcal{C}_{\alpha}}{3}$. When one knows that  the input states come from a polar cap, with maximal latitude $\theta_0$, the above fidelity can be achieved through a lower entangled state $\mathcal{C}_{\alpha^{\prime}}$ which from Eq. \eqref{eq:fidalpha} is computed to be
\begin{eqnarray}
\mathcal{C}_{\alpha^{\prime}} = \max \Big\lbrace 0, 1 - \frac{2(1-\mathcal{C}_{\alpha})}{(2+\cos \theta_0)(1-\cos\theta_0)} \Big \rbrace.
\label{eq:entlow}
\end{eqnarray}
For low enough $\theta_0$ values, we can, in principle, get $ \frac{2 + \mathcal{C}_{\alpha}}{3} \leq F_{cl}(\theta_0)$ by using Eq. \eqref{eq:fidclassical}. It implies that in presence of sufficient information about the inputs, the entanglement-free (classical) protocol is sufficient to reproduce the desired fidelity, as shown in Fig. \ref{fig:resource}. The maximization in Eq. \eqref{eq:entlow} emerges for this reason.  Similar analysis for the von Mises-Fisher density yields
\begin{eqnarray}
\mathcal{C}_{\alpha^{\prime}} = \max \Big\lbrace 0, 1 - \frac{2(1-\mathcal{C}_{\alpha})\kappa^2}{(1-\kappa \coth \kappa)} \Big \rbrace,
\label{eq:entlowfisher}
\end{eqnarray}
which also produces the similar advantage with \(1/\kappa\).

\subsubsection{Reduction in classical communication with prior information}

We now investigate how prior knowledge can alter the requirement of classical communication in the protocol. 
For a given shared state,  $|\psi_{\alpha}\rangle$ with non-maximal entanglement content, i.e., with $(\alpha < 1/2)$, the  probabilities of obtaining each Bell states after measurement  for a particular input state with Bloch angles $\theta$ and $\phi$ are given by
\begin{eqnarray}
p(\phi^{\pm}) = \frac{1}{2}(\alpha \cos^2 \theta/2 + (1-\alpha)\sin^2 \theta/2), \nonumber \\
p(\psi^{\pm}) = \frac{1}{2}(\alpha \sin^2 \theta/2 + (1-\alpha)\cos^2 \theta/2).
\end{eqnarray}
As expected, the azimuthal symmetry ($\phi$-independence) is reflected in the probabilities and \(p_i\)s also depend on the inputs.
However, such biasedness  does not allow one to reduce the required amount of classical communication since the average clicking probabilities when states are sampled uniformly from the Bloch sphere becomes identical, i.e.,
\begin{eqnarray}
\frac12 \int_0^\pi d\theta \sin \theta ~p(\phi^{\pm}) = \frac12 \int_0^\pi d\theta \sin \theta ~p(\psi^{\pm})= \frac14.  
\end{eqnarray}
Therefore, $2$-bits of CC is essential in the uniform scenario irrespective of the entanglement of the shared state. 
\begin{figure}[ht]
\includegraphics[width=0.7\linewidth]{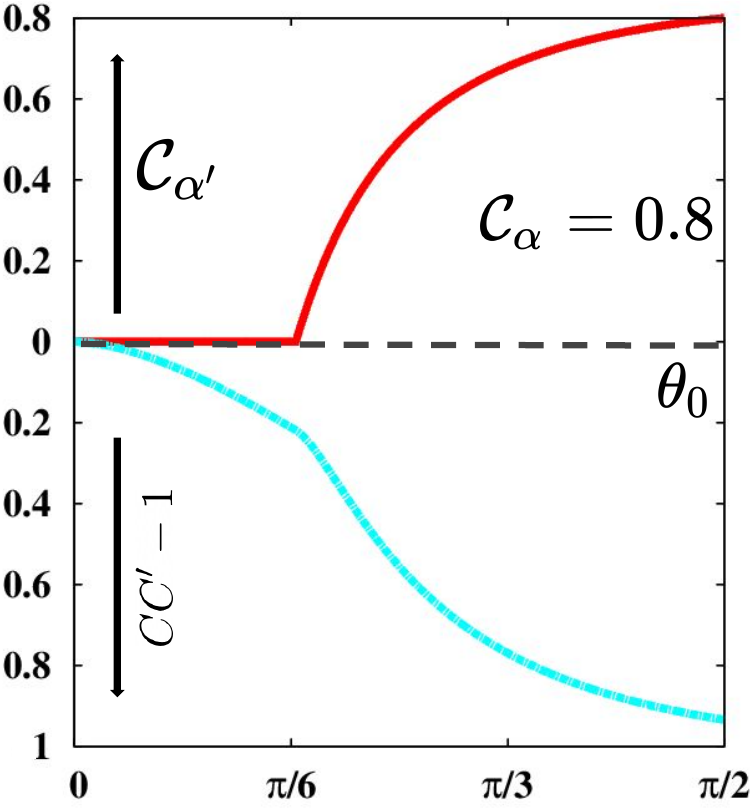}
\caption{Plot of  classical communication (CC') in bits, specifically \(CC'-1\) (lower panel) and entanglement in ebits ($\mathcal{C}_{\alpha'}$) (ordinate)  (upper panel) required to reproduce the same fidelity when a pure nonmaximally entangled with entanglement $\mathcal{C}_{\alpha} = 0.8$  with respect to the latitudinal extent of the polar cap $\theta_0$ (abscissa). The horizontal axis is dimensionless.}
\label{fig:ccgen}
\end{figure}

\begin{figure*}[ht]
\includegraphics[width=0.8\linewidth]{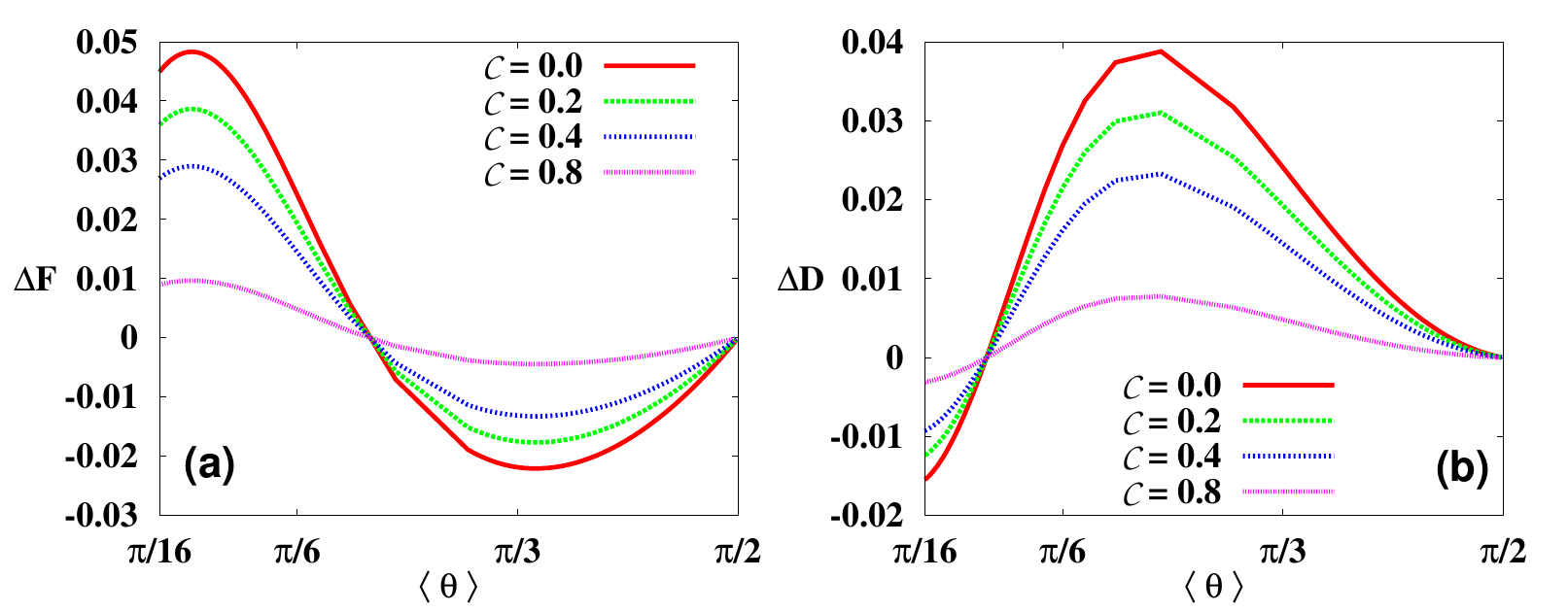}
\caption{ (a): Difference between average fidelities obtained by using polar cap and von Mises-Fisher distribution, \(\Delta F(t_1, t_2, t_3) \) for arbitrary two-qubit pure states (ordinate) against  fixed average polar angle, $\langle \theta \rangle$  (abscissae) obtained via Eq. (\ref{eq:avgpolar}). (b) Similar plot for fidelity deviation \(\Delta D(t_1, t_2, t_3) \). Both the axes are dimensionless. }
\label{fig:comp1}
\end{figure*}

In contrast to uniformly distributed inputs, if the state to be teleported is chosen from a non-uniform distribution, like the polar cap, we will now show that the biased clicking probabilities imply a reduced CC requirement provided classical   information transmission about the measurement outcomes from the sender to the receiver occurs via the noiseless channel. For a polar cap of latitude extent $\theta_0$, the outcomes of Bell measurement  on an average obtain with probabilities
\begin{eqnarray}
p(\phi^{\pm}) &=& \frac18 \lbrace \alpha (3 + \cos \theta_0) + (1-\alpha)(1 - \cos \theta_0) \rbrace, \nonumber \\
p(\psi^{\pm}) &=& \frac12 (1 - 2p(\phi^{\pm})),
\label{eq:probpolar}
\end{eqnarray} 
 while for the von Mises-Fisher density with a concentration parameter $\kappa$, the corresponding average probabilities read as
\begin{eqnarray}
p(\phi^{\pm}) &=& \frac{1}{4\kappa} \lbrace \kappa - (1-2\alpha)(\kappa \coth \kappa -1) \rbrace, \nonumber \\
p(\psi^{\pm}) &=& \frac12 (1 - 2p(\phi^{\pm})).
\label{eq:probfisher}
\end{eqnarray}
The amount of classical information transmission from Alice to Bob  can then be  quantified by evaluating  Shannon entropy of the probability distribution for obtaining the outcomes in Bell measurement, i.e., $H(\mathbf{X})$, with $\mathbf{X} = \lbrace p(\phi^+), p(\phi^-), p(\psi^+), p(\psi^-) \rbrace$, where $H(\lbrace p_i\rbrace) = -\sum_i p_i \log_2 p_i$. For consistency, when $\alpha = 1/2$, all the outcomes in the Bell measurement become equiprobable for any distribution of the input, thereby requiring $2$ bits of CC. However, for $\alpha < 1/2$, the average probabilities of obtaining outcomes differ from the uniform case and, therefore, one can faithfully transmit  \emph{classical information} of the measurement outcomes using less than $2$ bits of CC. 
In the limit of $\alpha \to 0$ and $\theta_0 \to 0$, we get the most asymmetric case where two of the Bell states click with probability $= 1/2$ while the other two does not click at all. It corresponds to  a $1$-bit of CC. The important feature here is that the biases of Bell clickings decrease as $\mathcal{C}_{\alpha}$  increases, and when the shared state is the maximally entangled state, we arrive at the completely symmetric case of equal probability for obtaining any Bell measurement outcomes with all states of the Bloch sphere, thereby requiring the full $2$-bits of CC.
Specifically, as shown  in Fig. \ref{fig:resource-cc}, we observe that for a fixed amount of entanglement in the shared state,  the amount of CC  decreases with the increase of  information available for inputs, i.e., with \(\theta_0\) and \(1/\kappa\). Interestingly, unlike the uniform case, the requirement of CC also depends on the shared entanglement.

\subsubsection{Prior information vs. entanglement vs. classical communication}

Let us now analyze the most general scenario where all the three players, namely prior information about the input, the entanglement content of the resource and the classical communication required for faithful communication of Bell measurement results are considered. In particular, for a given average fidelity, we analyze how the requirements of entanglement and CC cost decrease when some prior information of the input is available.

Let us first examine the situation when the inputs come from a polar cap distribution and $|\psi_{\alpha}\rangle$ is the resource state. For a given fidelity requirement, say $F$, when inputs come from an uniform distribution, one needs an entanglement $\mathcal{C}_{\alpha}= 3F - 2$ and $2$-bits of CC. When prior information via $\theta_0 < \pi/2$ is provided, the 
entanglement required to attain a fidelity $F$ is $\mathcal{C}_{\alpha^{\prime}}$ and is given in Eq. \eqref{eq:entlow}  while the CC requirement is $CC^{\prime} = H(\mathbf{X})$, where $\mathbf{X}$ is set of probabilities in obtaining Bell measurement outcomes given in Eq. \eqref{eq:probpolar} computed for the state parameter $\alpha^{\prime}$ and the latitudinal extent of the polar cap $\theta_0$.  Therefore, for a fixed fidelity, $F$, one can write
\begin{eqnarray}
\lbrace \theta_0 = \pi, \mathcal{C}_{\alpha}, CC = 2 \rbrace \equiv \lbrace \theta_0 < \pi/2, \mathcal{C}_{\alpha}^{\prime} < \mathcal{C}_{\alpha}, CC^{\prime} < 2 \rbrace. \nonumber \\
\end{eqnarray}
 We now plot $\mathcal{C}_{\alpha}^{\prime}$ and $CC^{\prime}$ with respect to $\theta_0$ to visualize the effect of decreasing resource consumption on increasing the prior information for both polar cap distributions, see Fig. \ref{fig:ccgen}.
We notice that once $\theta_0$ is increased, the CC requirement (CC') strictly increases while the required entanglement remains zero upto a critical value of $\theta_0 = \bar{\theta}_0$ before increasing monotonically with $\theta_0$. This implies that when the amount of prior information is high $\theta_0 < \bar{\theta}_0$, the classical (entanglement free) protocol can match the fidelity obtained in the case for uniform inputs using an entangled shared state.
  One can construct a similar relation for the von Mises-Fisher distribution also using Eqs. \eqref{eq:entlowfisher} and \eqref{eq:probfisher} which naturally show similar trends.
\begin{figure}[h]
\includegraphics[width=0.8\linewidth]{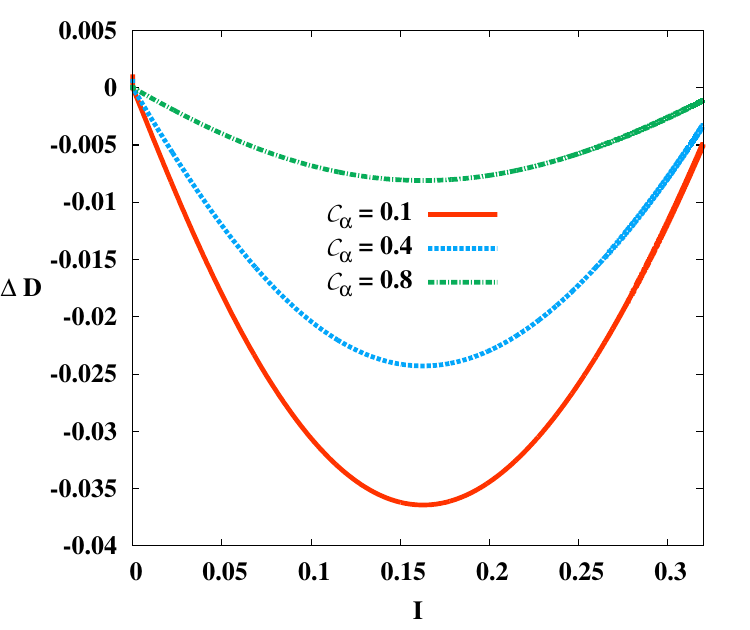}
\caption{Difference between the fidelity deviation for the polar cap  and von Mises-Fisher distributions,  \(\Delta D(t_1, t_2, t_3) \) (in \(y\)-axis) with respect to the classical information content as measured by $I$ (in \(x\)-axis) as in Eq. \eqref{eq:info1}. Different curves correspond to 
different values of entanglement  in the shared pure states. The deviation for the von Mises-Fisher density is universally greater than that of the polar cap value, thereby indicating advantage in choosing  polar cap distribution for inputs. Both the axes are dimensionless. }
\label{fig:comp2}
\end{figure}

\section{Comparative study between polar cap and Fisher density}
\label{sec:comparison}

When prior information about the input is available in the form of the two considered distributions, it is natural to compare their performance  in terms of their average fidelity yield and the corresponding fidelity deviation. To make a fair comparison, we adopt two methods. 

\emph{Based on average polar angle.} Let us choose inputs according to polar cap and von Mises-Fisher distributions with a fixed average polar angle. We achieve this by choosing $\theta_0 = \theta_0^*$ and $\kappa=\kappa^*$ in such a way that
\begin{eqnarray}
\langle \theta \rangle = \int \theta\sin \theta ~p_{\theta_0^*}(\theta,\phi) ~d\theta d\phi = \int \theta\sin \theta ~p_{\kappa^*}(\theta,\phi) ~d\theta d\phi. \nonumber \\
\label{eq:avgpolar}
\end{eqnarray} 
For a fixed average polar angle, $\langle \theta \rangle$,  $\theta_0^*$ and $\kappa^*$ values are obtained by inverting Eq. \eqref{eq:avgpolar}.
Now we compute and contrast both the fidelity and deviation for the polar cap and von Mises-Fisher distributions with $\theta_0 = \theta_0^*$ and $\kappa=\kappa^*$. 

Using Eqs. \eqref{eq:fidgenpolar} and \eqref{eq:test}, for a general shared state with $t_1, t_2,$ and $t_3$ being the elements of the diagonal correlation matrix,  we compute the difference between fidelity and deviation obtained via two distributions, i.e., 
\begin{eqnarray}
\Delta F(t_1, t_2, t_3) &=& F(\kappa^*) - F(\theta_0^*), \nonumber \\
\Delta D(t_1, t_2, t_3) &=& D(\kappa^*) - D(\theta_0^*).
\end{eqnarray}
For the case of non-maximally entangled pure states, we explicitly plot \(\Delta F(t_1, t_2, t_3)\)   and \(\Delta D(t_1, t_2, t_3)\)   in Fig. \ref{fig:comp1}. Between the two distributions with the same average polar angle, there is no ubiquitous one which yields a better performance in quantum teleportation.  In particular, we observe that for small values of  $\langle \theta \rangle$, polar cap performs better than von Mises-Fisher distribution for average fidelity while the opposite picture emerges for deviation as one expects from the definition. Interestingly, however, the value of $\langle \theta \rangle$ where von Mises-Fisher ditribution performs  better in case of average fidelity and worse for deviations occurs at different point, thereby displaying their nontrivial nature.

\emph{Based on prior information content.} Instead of the average polar angle, let us now consider the amount of information that one can  extract in the classical protocol as the common platform to compare the two chosen distributions on the Bloch sphere. In this case, for arbitrary pure two-qubit state in Eq. \eqref{eq:psialpha} considered before, a qualitatively different feature emerges. As discussed before, the maximal information that can be extracted classically is given by $F_{cl}$. Let us choose the values of $\theta_0^*$ and $\kappa^*$ which yields the same value of $F_{cl}$, i.e., $F_{cl}(\theta) = F_{cl}(\kappa)$. From Eqs. \eqref{eq:fidclassical} and \eqref{eq:fidclassicalfisher}, it leads to the  criterion,
\begin{eqnarray}
\frac{(2 + \cos \theta_0^*)(1 - \cos \theta_0^*)}{6} = \frac{\kappa^* \coth \kappa^* - 1}{(\kappa^*)^2}.
\label{eq:criterion-comp1}
\end{eqnarray}
For polar cap and von Mises-fisher distributions with these choices of $\theta_0$ and $\kappa$ values, the average fidelity for arbitrary two-qubit pure state  having a fixed value of entanglement turns out to be identical which can be seen by comparing  Eqs. \eqref{eq:fidalpha}, \eqref{eq:fidpsifisher}, and \eqref{eq:criterion-comp1}. However, we find that the fidelity deviation for the polar cap distribution is always smaller than that obtained via von Mises-Fisher density, as is clearly visible from the quantity  \(\Delta D(t_1, t_2, t_3)\) plotted in Fig. \ref{fig:comp2}.
Therefore, although both the distributions offer the same average fidelity, the polar cap distribution, on virtue of a smaller deviation count, can be considered as \emph{better} in comparison to the von Mises-Fisher distribution for choosing inputs  in quantum teleportation. 

Like in the preceding sections, comparative teleportation performances can also be studied for the family of BD states. For both the measures used for comparison, average value of $\theta$, and $F_{cl}$, the gap between the fidelity and deviation between the polar cap and von Mises-Fisher densities vanish for the Werner states. As a matter of fact, the gap is vanishing for any moments of fidelity, which follows directly from Eq. \eqref{eq:fidW1}. For other BD states, we get a varying value for this gap that varies from zero (the Werner state case) to that obtained for the nonmaximally entangled pure states which are identical to the BD states with rank-2.

\begin{figure}[ht]
\includegraphics[width=\linewidth]{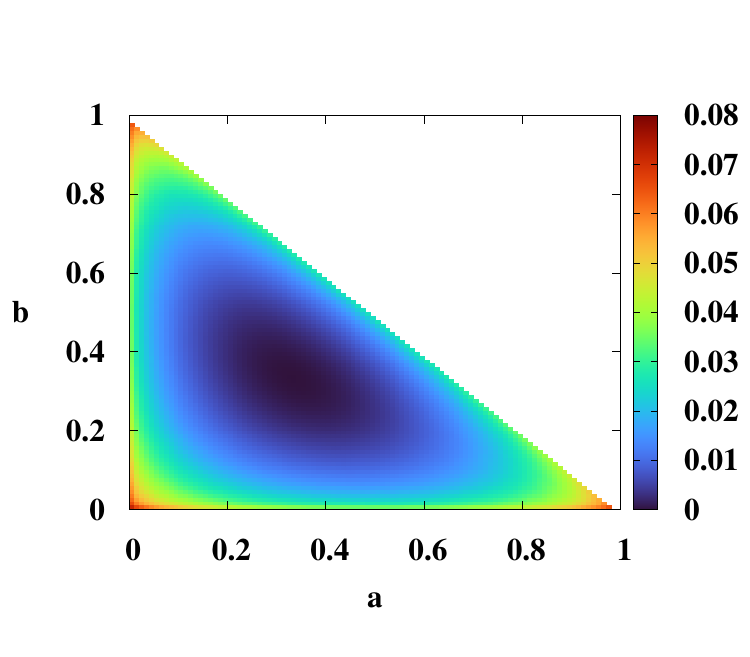}
\caption{Map plot of $\Delta F(\pi/4) = F(\pi/4) - F(\pi)$ in the plane of the Schmidt coefficients $a$  (horizontal axis) and $b$ (vertical axis)  of a non-maximally entangled pure qutrit  state in Eq. \eqref{eq:state3l}. $\Delta F(\pi/4) \geq 0$ in the entire $a-b$ plane conforms the fact that prior knowledge of the inputs leads to the enhancement in fidelity. All axes are dimensionless.}
\label{fig:3l1}
\end{figure}

\section{Role of non-uniform distribution in inputs in higher dimensional teleportation}
\label{sec:highdimen}

In this section, we go beyond two-qubits  \cite{highdim1, highdim2, ZeilingerPan'19} and investigate the role of distributions for the choices of inputs in the performance of teleportation. 
Let us consider an arbitrary two-qutrit shared pure state, given by
\begin{eqnarray}
|\psi\rangle = \sqrt{a}|00\rangle + \sqrt{b}|11\rangle + \sqrt{1-a-b}|22\rangle.
\label{eq:state3l}
\end{eqnarray}
Following the standard protocol \cite{ZeilingerPan'19}, the fidelity yield for an arbitrary input qutrit state $|\Phi\rangle = x |0\rangle + y |1\rangle + z |1\rangle$, where $|x|^2 + |y|^2 + |z|^2 = 1$, is computed as
\begin{eqnarray}
f(x,y,z) = |x|^4 + |y|^4 + |z|^4 &+& 2\big(\sqrt{a}(\sqrt{b}+\sqrt{1-a-b}) + \nonumber \\
 \sqrt{b(1-a-b)}~\big)\big(|x|^2|y|^2 &+& |z|^2(|x|^2 + |y|^2)\big).
 \label{eq:fid3l}
\end{eqnarray}
To compute the average fidelity, we parameterize $x = x_1 + ix_2, y = y_1 + iy_2,$ and $z = z_1 + iz_2$, where $i = \sqrt{-1}$ in the following way:
\begin{eqnarray}
x_1 &=& \cos \phi \sin \theta_1 \sin \theta_2 \sin \theta_3 \sin \theta_4 \nonumber \\
x_2 &=& \sin \phi \sin \theta_1 \sin \theta_2 \sin \theta_3 \sin \theta_4 \nonumber \\
y_1 &=& \cos \theta_1  \sin \theta_2 \sin \theta_3 \sin \theta_4 \nonumber \\
y_2 &=& \cos \theta_2 \sin \theta_3 \sin \theta_4 \nonumber \\
z_1 &=& \cos \theta_3 \sin \theta_4 \nonumber \\
z_2 &=& \cos \theta_4.
\label{eq:param}
\end{eqnarray}
To extract a Haar uniform average of $f(x,y,z)$ in Eq. \eqref{eq:fid3l} via the parameterization in Eq. \eqref{eq:param}, we integrate it using the measure $d \Omega = d\phi d\theta_1 d\theta_2 d\theta_3 d\theta_4 \sin \theta_1 \sin^2 \theta_2 \sin^3 \theta_3 \sin^4 \theta_4$ with a normalization constant of $\pi^3$, which turns out to be the volume of $\mathcal{S}^5$.
 It is obtained by integrating the measure $d\Omega$, where $0 \leq \theta_i \leq \pi$, and $0 \leq \phi \leq 2\pi$.
 
 We impose the prior information about the inputs by restricting the latitudinal extent, i.e., by constraining  one of the polar angles in Eq. \eqref{eq:param}, say $\theta_4$ upto $\theta_0 < \pi$. The updated normalization constant, $V(\theta_0)$ is computed by integrating $d \Omega$ where all the angles are integrated over their usual domain, excepting $\theta_4$, which runs from $0$ to $\theta_0$.
Under the same limits, we compute 
\begin{eqnarray}
F(\theta_0) = \frac{1}{V(\theta_0)}\int d\Omega ~f(x,y,z).
\end{eqnarray}
Like in earlier sections, we find that prior information about the distribution leads to fidelity enhancement, i.e., $F(\theta_0) \geq F(\pi)$. For a representative example, we choose $\theta_0 = \pi/4$ and plot $\Delta F(\pi/4) = F(\pi/4) - F(\pi)$ in Fig. \ref{fig:3l1}. In general, $\Delta F(\theta_0 < \pi) \geq 0$ demonstrates how prior information about the input state to be teleported leads to the enhancement of teleportation fidelity. Furthermore, when CC, entanglement and prior information are compared simultaneously, we get qualitatively similar features as obtained in the case of qubits in Fig. \ref{fig:ccgen}.

\subsection{Dimensional advantage in qutrit teleportation}

For qutrit teleportation, the classical fidelity for the uniform distribution, \(\frac{2}{d+1}\),  reduces to $1/2$. In presence of prior information of the inputs, the classical protocol, as in the case qubit teleportation, gives a higher fidelity than that of the uniform case. For the representative example of $\theta_4 \leq \pi/4$ in the qutrit scenario, the prior information content about the input states, from Eq. \eqref{eq:info2},  turns out to be $I_f \sim 0.16$. If we now consider the case of the polar cap distribution for qubit teleportation with the same value of $I_f$, we require an approximate $\theta_0$ value of $1.112$.

 In this configuration of a fixed  $I_f$ values in both the cases of qubit and qutrit teleportation schemes, we compute the mean of the average teleportation fidelity $\langle F \rangle_d$ for Haar uniformly generated pure shared states. 
We find  the percentage of enhancement of the obtained mean fidelity as computed by $$\eta_d = \Big(\frac{\langle F \rangle_d - F_{cl}}{F_{cl}}\Big)\times100.$$ 
We find that in the qubit case, $\eta_2 \sim 23 \%$, while for the qutrits,  $\eta_3 \sim 57 \%$. The larger enhancement for the qutrit case is observed for other values of $I_f$ as well. This feature is a strong signature of a dimensional improvement in teleportation protocol when the information about the distribution of the input states is available.

\section{Conclusion}
\label{sec:conclu}

Successful implementation of quantum teleportation protocol of unknown quantum states requires resources that include  shared entangled states, Bell measurement, and  classical communication from the sender to the receiver via a noiseless channel. In our paper, we altered the distribution of inputs from uniform to  non-uniform ones  and investigated the consequences of the performance of teleportation in terms of average fidelity and the second moment of fidelity, fidelity deviation. Specifically, we considered that the inputs are  chosen either from the polar cap or from the von Mises-Fisher distributions.

				In both the scenarios, we found analytical forms of average fidelities and their deviations in terms of correlators for arbitrary two-qubit states by following the standard teleportation scheme. Exploiting these formulae, we showed that these non-uniform distributions always give advantages for pure two-qubit states over the uniform ones. Similar advantages are also obtained for a class of Bell diagonal (BD) states and a mixture of nonmaxiamlly entangled states with product ones, although there exists a set of BD states which do not lead to any increments in fidelities for these choices of distributions, thereby indicating standard teleportation protocol to be sub-optimal for these classes of states. 
				
The effect of distributions for inputs on other resources necessary for quantum teleportation are also studied. In particular, we established a connection between	 the entanglement required between the sender and the receiver  for the polar cap as well as von Mises-Fisher distributions and the one with uniform distributions. Importantly, we found that the amount of required classical information communicated from the sender to the receiver decreases with the increase of information about the inputs. Such resource reduction in terms of classical information transfer following the standard teleportation scheme does not occur when the inputs are chosen from the uniform distribution. 

We also compared the performance  for both the non-uniform distributions with respect to average teleportation fidelities and deviations by using two kinds of figures of merits. In  the case of pure states, we found that for a given prior information, both the distributions yield the same average fidelity although the von Mises-Fisher distribution leads to a higher deviation compared to the polar cap ones, which  demonstrates the superiority of choosing polar cap distribution for inputs over the other one. We finally showed that a similar advantage can also be obtained if the shared state and inputs are taken from the higher dimensional systems. Moreover, we observed that the enhancement of average fidelity on average is higher for qutrits than that of  qubits, thereby showing the dimensional improvements even when prior information about inputs is available. Additionally, we showed that local operations and classical communication may sometimes be required based on the prior information about the inputs and shared states to increase the fidelity.

Our analysis revealed that prior information content has an important role to play in teleportation protocols.  Our work suggests that identifying resources to enhance the efficiencies of communication devices can be an interesting direction of further investigations.

\section*{Acknowledgement}
We acknowledge the support from Interdisciplinary Cyber Physical Systems (ICPS) program of the Department of Science and Technology (DST), India, Grant No.: DST/ICPS/QuST/Theme- 1/2019/23. Some numerical results have been obtained using the Quantum Information and Computation library (QIClib).  
This research was supported in part by the ‘INFOSYS scholarship for senior students.’ We also thank the anonymous 
Referees for valuable suggestions.


\begin{thebibliography}{100}

\bibitem{pirandola'15}  S. Pirandola, J. Eisert, C. Weedbrook, A. Furusawa, and S. L. Braunstein, Nat. Photonics {\bf 9}, 641 (2015).

\bibitem{bennett'93} C. Bennett, G. Brassard, C. Crepeau, R. Jozsa, A. Peres, and W. Wootters, Phys. Rev. Lett. {\bf 70}, 1895 (1993).

\bibitem{repeater'98} H.-J. Briegel, W. Du\"r, J. I. Cirac, and P. Zoller, Phys. Rev. Lett. {\bf 81}, 5932 (1998).
\bibitem{bennett'01} C. H. Bennett, D. P. DiVincenzo, P. W. Shor, J. A. Smolin, B. M. Terhal, and W. K. Wootters, Phys. Rev. Lett. {\bf 87}, 077902 (2001).
\bibitem{gottesman'99} D. Gottesman, and I. L. Chuang,  Nature {\bf 402}, 390 (1999).
\bibitem{gross'07} D. Gross, and J. Eisert, Phys. Rev. Lett. {\bf 98}, 220503 (2007).
\bibitem{murao'99} M. Murao, D. Jonathan, M. B. Plenio, and V. Vedral, Phys. Rev. A {\bf 59}, 156 (1999).
\bibitem{popescu'94} S. Popescu, Phys. Rev. Lett. {\bf 72}, 797 (1994).
\bibitem{jozsa'94} R. Jozsa,  J. Mod. Opt. 41, 2315 (1994).
\bibitem{horo'96}R. Horodecki, M. Horodecki and P. Horodecki, Phys. Lett. A {\bf 222} 21 (1996).

\bibitem{horo'99}  M. Horodecki, P. Horodecki, and R. Horodecki, Phys. Rev. A {\bf 60}, 1888 (1999).

\bibitem{MassarPop} S. Massar and S. Popescu, Phys. Rev. Lett. {\bf 74}, 1259 (1995).

\bibitem{caval'17} D.Cavalcanti, P.Skrzypczyk, and I.Supic, Phys.Rev.Lett. {\bf 119}, 110501 (2017).

\bibitem{bouwm'97}D. Bouwmeester, \emph{et al.}, Nature {\bf 390}, 575 (1997).
\bibitem{ursin'04} R. Ursin, \emph{et al.}, Nature {\bf 430}, 849 (2004).
\bibitem{ma'12}X. S. Ma,\emph{et al.}, Nature {\bf 489}, 269 (2012).
\bibitem{yin'12} J. Yin, \emph{et al.}, Nature {\bf 488}, 185 (2012).
\bibitem{bin'03} I. Marcikic, H. de Riedmatten, W. Tittel, H. Zbinden, and N. Gisin, Nature {\bf 421}, 509 (2003).
\bibitem{barrett'04}M. D. Barrett, \emph{et al.}, Nature {\bf 429}, 737 (2004).
\bibitem{nollek'13} C. N\"olleke, \emph{et al.}, Phys. Rev. Lett. {\bf 110}, 140403 (2013).
\bibitem{nmr} M. A. Nielsen, E. Knill, and R. Laflamme, Nature {\bf 396}, 52 (1998).
\bibitem{solid'13} L. Steffen, \emph{et al.},  Nature {\bf 500}, 319 (2013).
\bibitem{gao'13} W. B. Gao \emph{et al.},  Nature Commun. {\bf 4}, 2744 (2013).
\bibitem{tw} S. Roy, A. Bera, S. Mal, A. Sen De, U. Sen,  	arXiv:1905.04164 [quant-ph].
\bibitem{entanglementrev} R. Horodecki, P. Horodecki, M. Horodecki, and K. Horodecki,
Rev. Mod. Phys. {\bf 81}, 865 (2009). 


\bibitem{dist1} C. H. Bennett, G. Brassard, S. Popescu, B. Schumacher, J. Smolin, and W. K. Wooters, Phys. Rev. Lett. {\bf 76}, 722 (1996).
\bibitem{dist2} D. Deutsch, A Ekert, R. Jozsa, C. Macchiavello, S. Popescu, and A. Sanpera, Phys. Rev. Lett. {\bf 77}, 2818 (1996).
\bibitem{dist3} C. H. Bennett, D. P. Di Vincenzo, J. Smolin, and W. K. Wootters, Phys. Rev. A {\bf 54}, 3824 (1996).

\bibitem{dist4} M. Horodecki, P. Horodecki, and R. Horodecki, Phys. Rev. Lett. {\bf 78}, 574 (1997).

\bibitem{badziag'00} P. Badziag, M. Horodecki, P. Horodecki, and R. Horodecki, Phys. Rev. A {\bf 62}, 012311 (2000).

\bibitem{vers'03} F. Verstraete and H. Verschelde, Phys. Rev. Lett. {\bf 90}, 097901 (2003).
\bibitem{Li'20} J. Y Li, X. X. Fang, T. Zhang, G. N. M. Tabia, H. Lu, and Y.C. Liang1, arXiv: 2008.01689 (2020).

\bibitem{henderson'00} L. Henderson, L. Hardy, and V. Vedral, Phys. Rev. A {\bf 61}, 062306 (2000).
\bibitem{taketani'12} B. G. Taketani, F. de Melo, and R. L. de Matos Filho, Phys. Rev. A {\bf 85}, 020301(R) (2012).

 \bibitem{Fisher} R. Fisher, Proceedings of the Royal Society A: Mathematical, Physical and Engineering Sciences {\bf 217}, 295 (1953).



\bibitem{bang'18} J. Bang, J. Ryu, and D. Kaszlikowski,  J. Phys. A: Math. Theor. {\bf 51}, 135302 (2018).

\bibitem{ds} S. Das, A. Kumar, A. Sen De, U. Sen, 	arXiv:1903.03564 [quant-ph]. 

\bibitem{dev1} A. Ghosal, D. Das, S. Roy, and S. Bandyopadhyay, J. Phys. A: Math. Theor. {\bf 53}, 145304 (2020). 

\bibitem{arka'20} A. Ghosal, D. Das, S. Roy, and S. Bandyopadhyay, Phys. Rev. A {\bf 101}, 012304 (2020).

\bibitem{dev2} S. Roy and A. Ghosal, Phys. Rev. A {\bf 102}, 012428 (2020).

\bibitem{nielson'02} M. A. Nielsen, Phys. Lett. A {\bf 303}, 249 (2002).
\bibitem{gate'20} F. Wudarski,  J. Marshall,  A. Petukhov, and E. Rieffel, arXiv: 2006.03086 (2020).



 \bibitem{Cerf'02} N. J. Cerf, M. Bourennane, A. Karlsson, and N. Gisin, Phys. Rev. Lett. {\bf 88}, 127902 (2002); T. Durt, D. Kaszlikowski,  J.-L. Chen, and L. C. Kwek,
Phys. Rev. A {\bf 69}, 032313  (2004).

\bibitem{highdim3} L. Roa, A. Delgado, and I. Fuentes-Guridi,  Phys. Rev. A {\bf 68}, 022310 (2003).

\bibitem{Durt'08} T. Durt, D. Kaszlikowski, and L. C. Kwek, Phys. Rev. A {\bf 77}, 042318  (2008).
\bibitem{Vertesi'10} T. Vertesi, S. Pironio, and N. Brunner, Phys. Rev. Lett. {\bf 104}, 060401 (2010).

\bibitem{Cava'18}  P. Skrzypczyk and D. Cavalcanti, Phys. Rev. Lett. {\bf 120}, 260401 (2018).

\bibitem{highdim1} X.-M. Hu, C. Zhang, B.-H. Liu, Y.-F. Huang, C.-F. Li, and G.-C. Guo, 	arXiv:1904.12249. 

\bibitem{highdim2} A. Fonseca, Phys. Rev. A {\bf 100}, 062311  (2019).

\bibitem{ZeilingerPan'19} Y.-H. Luo, \emph{et. al.}, Phys. Rev. Lett. {\bf 123}, 070505  (2019).



\bibitem{Wootterseof} W. K. Wootters, Phys. Rev. Lett. {\bf 80}, 2245 (1998). 

\bibitem{Werner'89} R. F. Werner, Phys. Rev. A {\bf 40}, 4277  (1989).

\bibitem{BD} The Bell diagonal states can be written as \(\rho= \sum_i p_i |\psi_i\rangle \langle \psi_i|\), where \(|\psi_i \rangle = |\phi^{\pm}\rangle, |\psi^{\pm}\rangle\), with \(|\phi^{\pm} \rangle = \frac{1}{\sqrt{2}}(|00\rangle \pm |11\rangle) \) and \(|\psi^{\pm} \rangle = \frac{1}{\sqrt{2}}(|01\rangle \pm |10\rangle) \). Depending on the upper bound on the sum, the rank of the state ranges from one to four. 


\bibitem{book} N.I.  Fisher,  T. Lewis, and B.J.J. Embleton, T.  Fisher, \emph{Statistical analysis of spherical data} (Cambridge University Press (1987)).



\end{thebibliography}
\end{document}